\documentstyle[aps,multicol]{revtex}
\input{epsf}
\begin{document}
\title{Security of quantum key distribution with imperfect devices}
\author{Daniel Gottesman,$^1$ Hoi-Kwong Lo,$^2$, Norbert L\"utkenhaus,$^3$ and John Preskill$^4$}
\address{$^1$ Perimeter Institute, Waterloo ON N2V 1Z3, Canada\\
$^2$ Department of Electrical and Computer Engineering and Department of Physics, \\University of Toronto, Toronto, Canada M5G 3G4\\
$^3$ Institute of Theoretical Physics and Max Planck Research Group,
Institute of Optics, Information, and Photonics,\\University of Erlangen--Nuremberg, 91058 Erlangen, Germany\\
$^4$ Institute for Quantum Information, California Institute of Technology, 
Pasadena, CA 91125, USA}
\maketitle
\begin{abstract}

We prove the security of the Bennett-Brassard (BB84) quantum key distribution protocol in the case where the source and detector are under the limited control of an adversary. Our proof applies when both the source and the detector have small basis-dependent flaws, as is typical in practical implementations of the protocol. We derive a general lower bound on the asymptotic key generation rate for weakly basis-dependent eavesdropping attacks, and also estimate the rate in some special cases: sources that emit weak coherent states with random phases, detectors with basis-dependent efficiency, and misaligned sources and detectors.

\end{abstract}
\pacs{PACS numbers: 03.67.Dd}
 
\begin{multicols}{2}

\section{Introduction}

The security of quantum cryptography is founded on principles of fundamental physics, rather than assumptions about the resources available to a potential adversary. In the BB84 quantum key distribution protocol \cite{BennettBrassard84}, two parties (Alice and Bob) establish a secret key about which the eavesdropper (Eve) cannot obtain a significant amount of information. Alice sends a key bit to Bob by preparing a qubit in one of two conjugate bases and Bob measures the qubit in one of the two bases; Eve, who does not know the basis chosen by Alice or by Bob, cannot collect information about the key without producing a detectable disturbance. This protocol, when suitably augmented by classical error correction and privacy amplification, is provably secure against any attack by Eve allowed by quantum mechanics \cite{Mayers96,LoChau99,BBBMR00,ShorPreskill00,KoashiPreskill02}.

Though security can be proven without imposing any restriction on Eve's attack (other than the requirement that she has no {\it a priori} information about the basis used), it is necessary to place conditions on the performance of the source and detector employed in the protocol. In the Shor-Preskill proof \cite{ShorPreskill00}, it is assumed that any flaws in the source and detector can be absorbed into Eve's basis-independent attack. In the proof by Mayers \cite{Mayers96}, the source is assumed to be perfect, but the detector is completely uncharacterized. In the Koashi-Preskill proof \cite{KoashiPreskill02}, the detector is perfect, but the source is uncharacterized, aside from the proviso that it leaks no information about the basis choice to Eve. In all of these cases, serious faults in the apparatus can be detected in the protocol, so that Alice and Bob will reject the key if the equipment performs badly. 

But none of these proofs apply when both the source and detector have small imperfections that depend on the basis used in the protocol, the case relevant to typical real-world implementations of quantum key distribution. Since the BB84 protocol with perfect sources and detectors is secure, it is intuitively clear that BB84 should remain secure if the imperfections are ``sufficiently small.'' We will sharpen this intuition into a quantitative statement, by calculating how the rate of generation of private key depends on the tolerance to which the equipment is characterized. 

The simplest way to analyze the consequences of characterized imperfections is to absorb the defective performance of the equipment into the eavesdropper's attack. Primarily for this reason, we are led to consider the security of the BB84 protocol in a different framework than in previous security proofs: the flaws in the source and detector may depend on the bases chosen, and furthermore Eve may know these bases, but her power to exploit this knowledge is limited. We will prove security under an {\em assumption} that limits the basis dependence of Eve's attack. 

It is natural to ask whether this assumption can be verified by conducting suitable tests on the source and detector (perhaps with testing equipment that is also not fully trustworthy, as in \cite{MayersYao98}). For now we put aside the issue of testing the equipment, and we will trust that our equipment performs {\em approximately} as expected. However, as cautious cryptologists we will assume that, within the prescribed limits, the performance of the equipment is controlled by Eve and maximally exploited by her to gain information about the shared key.

Our analysis follows the method of Shor and Preskill \cite{ShorPreskill00}, who proved the security of BB84 by relating it to an entanglement distillation protocol. Their argument exploited a symmetry between the two bases used in the protocol, whose consequence is that the entangled pairs to be distilled have equal rates of bit errors and phase errors. Our task in this paper is to extend the analysis to the case where the symmetry between the bases is broken because the adversary has information about the basis used. We will give a general argument showing that if the basis-dependence of the attack is sufficiently weak, then the gap between the bit error rate and the phase error rate is small; this argument allows us to establish security against arbitrary attacks that satisfy a particular criterion for weak basis dependence, and to derive a lower bound on the asymptotic key generation rate. 

To formulate our criterion for the attack to be weakly basis dependent, we focus on the coin that is flipped to determine the basis --- the basis dependence is weak if the adversary interacts only weakly with the coin. The Shor-Preskill argument shows that, for the purpose of analyzing the security of BB84, it is convenient to imagine that Alice chooses each of her key bits by measuring half of an entangled state, and that she delays these measurements until after Eve has attacked the signals. Likewise, for analyzing weakly basis-dependent attacks, we find it convenient to imagine that the coin flip that determines the basis is realized by measuring a qubit, and that this measurement is delayed until after the adversary's attack. Then
we can quantify the extent of the adversary's interactions with the coin according to how much the state of the coin is {\em disturbed}. Our general argument shows that if the disturbance of the coin is slight, then a secure key can be generated at a calculable nonzero rate.

Aside from presenting this general argument, we will also apply our methods to a few specific scenarios in which quantum key distribution is executed with imperfect devices. In some of these special cases, we can derive tighter lower bounds on the key generation rate than are obtained by the general argument. The examples we discuss include:
\begin{description}

\item{} {\em Tagging}. A faulty source may ``tag'' some of the qubits with information, readable by the eavesdropper, that reveals the basis used in the preparation. An important special case, also recently analyzed by Inamori, L\"utkenhaus and Mayers \cite{ILM01}, is a source emitting weak coherent states which with nonnegligible probability contain multiple photons prepared in the same polarization state. An adversary might intercept the extra photons and collect information about the basis used without causing any disturbance, compromising security.

\item{} {\em Basis-dependent detector efficiency}. If the detector sometimes misfires, the probability that a qubit is successfully detected might depend on the basis used. An adversary that can control whether the detector fires can use this power to disguise eavesdropping.

\item{} {\em Basis-dependent misalignment in the source or detector}. The source or detector might not be properly aligned to emit or detect a qubit in the desired basis. The adversary can exploit her freedom to rotate these devices to reduce the disturbance caused by her eavesdropping.
\end{description}

Our results do not subsume, nor are they subsumed by, the results of \cite{Mayers96,KoashiPreskill02,ILM01}.
Mayers and Koashi-Preskill assume that the detector or source is uncharacterized, but that the adversary is unable to influence the behavior of the devices to suit her purposes. We assume that the flaws in the devices are limited, but that the adversary controls the apparatus within these limits; furthermore, our security proof (unlike the Koashi-Preskill proof) applies to a source that leaks a small amount of information about the choice of basis. And more important, while Mayers assumes that the source is perfect, and Koashi and Preskill assume that the detector is perfect, our new techniques apply when both the source and the detector have small basis-dependent imperfections, the generic case in practical settings. In addition, while Koashi and Preskill assume that the signals emitted by Alice's source are uncorrelated with one another (the state describing the emission of $n$ signals is a tensor product of $n$ individual signals), and while Mayers likewise assumes that the signals are detected individually rather than collectively by Bob, our results are not inherently subject to such limitations. (However, we {\em do} assume that the signals are emitted and detected individually in many of the examples that we analyze.)

Aspects of the security of quantum key distribution in realistic settings have been analyzed previously \cite{Slutsky98,Norbert99,Brassard99,Felix01,Gilbert00,Gilbert01}. However, our proof of security holds for arbitrary collective attacks by the eavesdropper, while individual attacks were considered  in most previous work. (An important exception is the recent study by Inamori, L\"utkenhaus and Mayers \cite{ILM01} of sources that emit weak coherent states.) Although our results do not yet constitute a definitive analysis of the security of realistic quantum cryptography, we expect that the tools we have developed will prove useful in further studies.

Beyond any of our particular results, we have broadened notably the domain of applicability of the Shor-Preskill method for proving security. This method has many further applications, and in particular allows one to easily analyze the effectiveness of various enhancements of the protocol such as two-way postprocessing \cite{GottesmanLo01}.

Our findings are of both practical and conceptual interest. It is important to address whether practical implementations of quantum key distribution are truly secure, and in real-world implementations the apparatus is never flawless. And apart from practical concerns, quantum key distribution provides a fascinating theoretical laboratory for quantitatively exploring the unavoidable tradeoff between collecting information about a quantum system and disturbing the system.

We note that the security against arbitrary eavesdropping attacks of quantum key distribution performed with imperfect devices has also been analyzed by Ben-Or \cite{Ben-Or}.

The rest of this paper is organized as follows. In Sec.~II, we clarify the setting of our analysis by introducing Eve's collaborator Fred, who controls the flaws in the source and detector. We review the connection between the BB84 protocol and entanglement-based quantum key distribution in Sec.~III, and reprise the Shor-Preskill argument which is the foundation for all that follows. We carefully formulate our models for sources and detectors in Sec.~IV, and point out in Sec.~V some ways in which these models fail to capture fully the properties of real devices. In Sec.~VI we introduce the concept of a quantum coin, which is a useful tool for analyzing the power of Fred's basis-dependent attack on the equipment, and in Sec.~VII we present our security proof for a general class of attacks that depend sufficiently weakly on the basis. We then proceed to explore various applications of this result: In Sec.~VIII, we prove security for the case where the detector is perfect but the source has small generic flaws, and in Sec.~IX, we treat the case where the detector has small flaws and the source is flawed but {\em oblivious}; that is, it leaks no information about the basis used. Sec.~X analyzes the case where the source and detector are both slightly misaligned, and in Sec.~XI we state without proof a result for the case where the both the source and detector have small generic flaws (where the source is not necessarily oblivious). The case in which a fraction of the signals emitted by the source are tagged with basis information is dealt with in Sec.~XII; this analysis is relevant to sources that emit weak coherent states with random phases, sources that are close to single photon sources, and a scenario where some of the basis and key bits are selected by flipping biased coins. Finally, in Sec.~XIII we discuss the case of a detector with imperfect efficiency that is controlled by the adversary, and Sec.~XIV contains some concluding comments.

\section{Alice and Bob and Eve and Fred}
To clarify our assumptions about the source and detector imperfections, it is helpful to imagine that two collaborating adversaries are trying to foil the key distribution protocol: Eve and Fred. The goal of Alice and Bob is to generate a shared key not known to the Eve/Fred alliance. 

Fred knows the basis chosen by Alice and/or Bob, and he can tamper with the source and/or detector, but only within certain prescribed limits. Because the basis dependence of his attack is limited, Fred can acquire only limited knowledge of what signal was emitted by the source and what outcome was recorded by the detector. 

Eve on the other hand has no {\em a priori} knowledge of the basis chosen by Alice or by Bob, and she has no direct control over the source or the detector. But Eve is permitted to attack all of the signals sent by Alice to Bob collectively in any manner allowed by quantum mechanics. For example, Eve may entangle an ancilla that she controls with each signal after the signal is emitted by the source and before it is absorbed by the detector. Then Eve may delay the measurement of her ancilla until after all public discussion by Alice and Bob is concluded, choosing her measurement to optimize her information about the key. 

While Eve can send to Fred any quantum or classical message of her choice, communication from Fred to Eve is restricted. Before Eve interacts with the signals, Fred may wish to notify her about Alice's basis choice, but his only means of conveying this information is through his limited ability to control the source. After Bob confirms receipt of the signals, Fred is permitted to share further information with Eve by sending it via a classical or quantum side channel. Apart from this restriction on their communication, Eve and Fred are free to choose a common strategy that optimally exploits Fred's limited power to manipulate the source and detector.

Various security proofs apply to settings that can be distinguished by describing Fred's role. In the setting considered by Mayers and by Koashi and Preskill, Fred does not share information with Eve, and the goal of Alice and Bob is to generate a shared key that Eve does not know. Mayers assumes that the source is perfect, but Fred is free to choose the measurement performed by the detector, which can depend at Fred's discretion on Bob's declared basis, and to report to Bob a portion of the information collected in the measurement. Koashi and Preskill assume that the detector is perfect, but Fred is free to choose the states emitted by the source except for one proviso: the emitted state, averaged over Alice's key bit, is independent of Alice's basis. In the setting considered in this paper, Eve again applies an arbitrary basis-independent quantum operation to her probe and the transmitted signals. And again, Fred, who has information about the declared bases, can influence how the equipment operates. But now, the basis dependence of Fred's attack is limited, and Fred and Eve can pool their knowledge {\em after} the signals are detected.

All of these settings are interesting. In the Mayers model, the detector can be arbitrarily flaky, and in the Koashi-Preskill model, the source leaks no basis information but is otherwise arbitrary. In the model we consider, both the source and detector are ``pretty good'' but controlled (within limits) by the adversary. Our models of the source and the detector will be described in more detail in Sec.~IV.

\section{Distillation and privacy}
\label{sec:shor-preskill}

Our analysis follows the method of proof used by Shor and Preskill \cite{ShorPreskill00}, which we will now briefly review. In this method, security is first established, following \cite{LoChau99}, for a protocol based on an entanglement distillation protocol (EDP). Then the security of a ``prepare-and-measure'' protocol, namely BB84, is established through a reduction to the EDP protocol. 

We remark that entanglement distillation was first discussed in \cite{Bennett95}, that its relevance to the security of quantum key distribution was emphasized in \cite{Deutsch95}, and that this connection was established rigorously in \cite{LoChau99}. Entanglement distillation protocols have also been called ``entanglement purification protocols,'' abbreviated EPP. We prefer to say ``distillation'' rather than ``purification'' as ``purification'' now has another widely accepted meaning in quantum information theory.

In the EDP protocol, Alice creates $n + m$ pairs of qubits, each in the state
\begin{equation}
|\phi^+\rangle= {1\over\sqrt{2}}\left(|00\rangle + |11\rangle\right)~,
\end{equation}
the simultaneous eigenstate with eigenvalue one of the two commuting operators $X \otimes X$ and $Z\otimes Z$, where
\begin{equation}
X=\pmatrix{0&1\cr 1&0\cr}~,\quad Z=\pmatrix{1&0\cr 0&-1\cr}
\end{equation}
are the Pauli operators. Then she sends half of each pair to Bob. Alice and Bob sacrifice $m$ randomly selected pairs to test the ``error rates'' in the $X$ and $Z$ bases by measuring $X\otimes X$ and $Z\otimes Z$. If the error rate is too high, they abort the protocol. Otherwise, they conduct the EDP, extracting $k$ high-fidelity pairs from the $n$ noisy pairs. Finally, Alice and Bob both measure $Z$ on each of these pairs, producing a $k$-bit shared random key about which Eve has negligible information. The protocol is secure because the EDP removes Eve's entanglement with the pairs, leaving her powerless to discern the outcome of the measurements by Alice and Bob.

If the EDP protocol has special properties, then proving the security of BB84 can be reduced to proving security of the EDP.  Shor and Preskill considered EDP's with one-way communication \cite{BDSW96}, which are equivalent to quantum error-correcting codes, and furthermore, considered the specific class of codes known as Calderbank-Shor-Steane (CSS) codes \cite{CalderbankShor96,Steane96}. (Gottesman and Lo \cite{GottesmanLo01} have described how a similar reduction can be applied to certain EDP's with two-way communication.) Like any quantum error-correcting code, a CSS code can correct both bit errors (pairs with $Z\otimes Z=-1$) and phase errors (pairs with $X\otimes X=-1$). But the crucial property of a CSS code is that the bit and phase error correction procedures can be decoupled --- $Z$ errors can be corrected without knowing anything about the $X$ errors and vice-versa. 

In the EDP protocol, the key is affected by the bit error correction but not by the phase error correction. The phase error correction is important to expunge entanglement with Eve and so ensure the privacy of the key. But Eve's information about the final key is unaffected if Alice and Bob dispense with the phase error correction. What is essential is not that the phase error correction is actually done, but rather that it would have been successful if it had been done.

With the phase error correction removed, the extraction of the final key from the $n$ noisy pairs is much simplified. Rather than first carrying out the EDP and then measuring $Z$ for each of the $k$ distilled pairs, Alice and Bob can instead measure $Z$ for each of the $n$ noisy pairs, and then do classical postprocessing of their measurement results to extract the final key. In this form, the entanglement-based protocol becomes equivalent to BB84. 

We can see the equivalence more clearly by adding one further wrinkle to the entanglement-based protocol. In the BB84 protocol, Alice and Bob choose their bases at random, so that about half of the sifted key bits are transmitted in the $X$ basis and about half in the $Z$ basis. But in the entanglement-based protocol as we have described it, all of the final key bits are generated by measuring in the $Z$ basis. To relate the two protocols, suppose that in the entanglement-based protocol a subset of the pairs is selected at random, and that for each pair in this subset, Alice and Bob apply the Hadamard transformation $H:X\leftrightarrow Z$ to their qubits before measuring $Z$. Equivalently, we can instruct Alice and Bob to measure $X$ rather than $Z$ for these selected qubits. Each measurement by Alice in the entanglement-based protocol prepares a qubit to be sent to Bob in one of the four BB84 states: $X=\pm 1,Z=\pm 1$, chosen at random. In BB84, Bob measures either $X$ or $Z$, and through public discussion Alice and Bob reject the key bits where they used different bases; the remaining key, for which their bases agree, is called the ``sifted key.'' As far as an eavesdropper is concerned, there is no difference between generating a bit of sifted key in BB84, where a qubit is prepared by Alice in a randomly chosen eigenstate of either $X$ or $Z$ and measured by Bob in the same basis, and generating a bit of key in the entanglement-based protocol, where Alice and Bob both measure their halves of an entangled pair of qubits. 

A vestige of the CSS code of the EDP survives as a scheme for error correction and privacy amplification in this prepare-and-measure protocol. In a CSS code, classical linear codes $C_1$ and $C_2^\perp$ are used for bit and phase error correction respectively, where $C_2\subset C_1$. The entanglement-based protocol is secure (whether or not the phase error correction is done) if, with ``high probability'' (probability of success exponentially close to unity), $C_1$ can correct the bit errors and $C_2^\perp$ can correct the phase errors. In the BB84 protocol, $C_1$ is used to correct bit errors in the key, and $C_2$ to amplify privacy. Specifically, Alice transmits the random string $w$ through the quantum channel, randomly selects a codeword $u$ of $C_1$, and announces $u+w$. Bob receives the corrupted string $w+e$, computes $u+e$, and corrects to $u$. The final key is the coset $u+ C_2$ of $C_2$ in $C_1$. 

If this method is used to compute the final key in the BB84 protocol, and if the key being distributed is very long, at what asymptotic rate can secure final key be extracted from the sifted key? The answer is the rate $k/n$ at which high-fidelity pairs can be distilled from noisy pairs in the EDP, which depends on how noisy the pairs are. The purpose of the verification test included in the protocol is to obtain a reliable estimate of the noise. In the EDP, a useful way to characterize the noise is to imagine that, after the final Hadamard transformations are applied to the pairs, all $n$ pairs are measured in the Bell basis --- that is, both $Z\otimes Z$ and $X\otimes X$ are measured. If there were no noise at all, we would find $Z\otimes Z=X\otimes X =1$ for every pair. Denote by $n\tilde\delta$ the number of pairs for which we have $Z\otimes Z=-1$ instead; we say that $\tilde\delta$ is the bit error rate of the noisy pairs. Denote by $n\tilde\delta_p$ the number of pairs for which we have $X\otimes X=-1$; we say that $\tilde\delta_p$ is the phase error rate of the pairs. 

For a given state of the $n$ pairs, the rates $\tilde\delta$ and $\tilde\delta_p$ are actually random variables, because the quantum measurement of the pairs is nondeterministic. But suppose that from the verification test, we can infer that for sufficiently large $n$ and any $\varepsilon > 0$, the inequalities $\tilde\delta < \delta +\varepsilon$ and $\tilde \delta_p< \delta_p +\varepsilon$ are satisfied with high probability. Furthermore, we may imagine that the key bits are subjected in the protocol to a publicly announced random permutation (or equivalently that the CSS code is correspondingly randomized), so that the bit and phase errors are randomly distributed among the qubits. It can then be shown \cite{GottesmanPreskill01,Hamada03} that, for sufficiently large $n$ and any $\varepsilon'> 0$, there exists a CSS code such that the EDP distills $k$ high-fidelity pairs from the $n$ noisy pairs, where
\begin{equation}
k/n>  1 - H_2(\delta+\varepsilon +\varepsilon') -H_2(\delta_p+\varepsilon +\varepsilon')~,
\end{equation}
and $H_2(\delta)= -\delta\log_2\delta - (1-\delta)\log_2(1-\delta)$ is the binary entropy function. Therefore, in BB84, we establish an asymptotically achievable rate of extraction of secure final key from sifted key (``key generation rate''):
\begin{equation}
\label{key_rate}
R= 1 - H_2(\delta) -H_2(\delta_p)~,
\end{equation}
That is, in the BB84 protocol, a fraction $H_2(\delta)$ of the sifted key bits are sacrificed asymptotically to perform error correction and a fraction $H_2(\delta_p)$ of the sifted key bits are sacrificed to perform privacy amplification. 

We note that, although the permutation randomizes the positions of both the bit errors and the phase errors, correlations between bit errors and phase errors may remain. However, these correlations do not affect the achievable rate, because with CSS codes the bit error correction and phase error correction are performed separately. We also remark that the code $C_1$ used to correct bit errors can be chosen to be efficiently decodable \cite{spielman}. It may not be possible to simultaneously choose the code $C_2^\perp$ to be efficiently decodable, but this is not important, since the phase error correction using $C_2^\perp$ is not actually carried out in the BB84 protocol --- it need only be possible in principle.

Our arguments so far have reduced the problem of demonstrating the security of BB84 to inferring sufficiently stringent upper bounds on both the bit error rate and the phase error rate of the pairs used to generate the key in the corresponding entanglement-based protocol, based on the results of the verification test. Inferring the upper bound on the bit error rate is straightforward. Let us consider the version of the entanglement-based protocol in which Alice and Bob measure both the test pairs and the key generating pairs in the $Z$ basis, but a Hadamard transformation is applied to randomly selected pairs just prior to the measurement. When Eve interacts with the qubits traveling from Alice to Bob, she has no {\em a priori} knowledge concerning which pairs will be used for the test and which will be used for key generation. Therefore, the test pairs are a fair sample; it follows from classical sampling theory that the joint probability of observing $n\delta$ errors in the test set and more than $n(\delta + \varepsilon)$ errors in the key set is exponentially small for any $\varepsilon > 0$ and $n$ sufficiently large. Note that this argument works even if Eve's attack induces strong correlations among the pairs; all that is required is that the sample selected for the test is chosen randomly.

Inferring an upper bound on the phase error rate requires an extra step. In the Shor-Preskill argument, it is assumed that the adversary has no {\em a priori} knowledge about the basis that Alice uses to send her signals and Bob uses to detect them. In the entanglement-based protocol, this becomes the statement that the adversary does not know to which pairs the Hadamard transformation is applied. But since the Hadamard interchanges the bit errors and the phase errors, it enforces a symmetry between the two types of errors. Therefore, the error rate measured in the test serves as an estimate of the phase error rate as well as the bit error rate: with high probability the phase error rate of the key generating pairs is also less than $\delta+\varepsilon$. We conclude that final key can be extracted from sifted key at the rate $R=1-2H_2(\delta)$.

Now we have sketched the complete proof of security of BB84, except for one technicality. The sampling theory argument actually shows that the {\em joint} probability of a error rate $\delta$ in the test pairs and an error greater than $\delta +\varepsilon$ for the key generating pairs is exponentially small. For a security analysis, we should show that the {\em conditional} probability of an error rate for the key generating pairs greater than $\delta+\varepsilon$ is exponentially small, given the error rate $\delta$ found in the test. The desired result follows from Bayes's theorem as long as we assume that Eve's attack ``passes'' the verification test with a probability that is not itself exponentially small. That is, we exclude strategies by Eve such that extraordinary luck is required to induce the (small) error rate $\delta$ found in the test.
With this caveat in mind, we propose this definition of security:

\medskip
\noindent{\bf Definition. Security of quantum key distribution}. {\em A quantum key distribution protocol is secure if for any attack by Eve that passes the verification test with a probability that is not exponentially small, with high probability Alice and Bob agree on a final key that is nearly uniformly distributed and Eve's information about the final key is exponentially small. Here ``exponentially small'' means bounded above by $e^{-CN}$ where $N$ is the number of signals transmitted in the protocol and C is a positive constant,  ``high probability'' means exponentially close to 1, and ``nearly uniformly distributed'' means with a probability distribution exponentially close to the uniform distribution. }
\medskip

\noindent And we conclude:

\medskip
\noindent{\bf Theorem 1. Security of BB84 against basis-independent attacks}. {\em The BB84 protocol is secure if Eve launches a basis-independent attack. Secure final key can be extracted from sifted key at the asymptotic rate
\begin{equation}
R={\rm Max}\big(1- 2H_2(\delta),0\big)
\end{equation}	  		                     
where $\delta$  is the bit error rate found in the verification test (assuming $\delta <1/2$).}
\medskip

To reiterate, two error rates are relevant to whether quantum key distribution is successful. The bit error rate is ``measured'' by conducting a verification test on a randomly sampled subset of the sifted key bits; that is, the observed bit error rate $\delta$ found in the test provides an estimate of the error rate $\tilde\delta$ in the key generating bits that is accurate with high probability. If $\delta$ is low enough, we can be confident that error correction will succeed, so that Alice and Bob share a common key. The phase error rate $\tilde\delta_p$ is not measured by direct sampling --- rather an upper bound $\tilde\delta_p < \delta_p+\varepsilon$  is inferred from the bit error rate. If the inferred phase error rate $\delta_p$ is low enough, we can be confident that phase error correction (if done) will succeed, so that Eve will have a negligible amount of information about the key.

If the adversary has no knowledge of the basis, then with high probability the gap $|\tilde\delta_p-\tilde\delta|$ between the bit and phase error rates is asymptotically negligible, and the inference is straightforward. For example, if the effect of Eve's attack is to apply $X$ to Bob's qubit, this action will induce a bit error if Alice and Bob both measure $Z$ to generate a key bit in the entanglement-based protocol, and it will induce a phase error if Alice and Bob both measure $X$. Since Eve doesn't know the basis, her action generates bit errors and phase errors with the same probability. But in this paper, going beyond Shor's and Preskill's original argument, we will allow Fred to know the basis, enabling him to enhance $\tilde\delta_p$ relative to $\tilde\delta$. In many cases of interest, the basis dependence of Fred's attack is limited; we can infer an upper bound $\tilde\delta_p < \delta_p +\varepsilon$ and so  through Eq.~(\ref{key_rate}) establish an achievable key length. 

In the BB84 protocol, Alice and Bob can measure both the error rate $\delta_X$ when they use the $X$ basis and the error rate $\delta_Z$ when they use the $Z$ basis. These rates need not be equal even if Eve does not know the bases that Alice and Bob use. For example, Eve might measure in the $Z$ basis each qubit she receives from Alice, and resend to Bob the $Z$ eigenstate found by her measurement, resulting in expected values $\delta_Z=0$ and $\delta_X=1/2$. We emphasize that  $\delta$ and $\delta_p$ should not be confused with $\delta_Z$ and $\delta_X$. The bit error rate $\delta\approx \left(\delta_X +\delta_Z\right)/2$ is observed in the verification test, but the phase error rate $\delta_p$ is not directly ``measured'' in the protocol.\footnote{If Alice and Bob perform a {\em refined error analysis} \cite{LoChauArdehali00} (measuring separate error rates for the two bases), they can improve the key generation rate to $R=1-H_2(\delta_Z)-H_2(\delta_X)$ \cite{Hamada03}.}

\section{Model devices}
\label{sec:model}
Because the Shor-Preskill argument, both in its original incarnation and in its extension to basis-dependent attacks, makes use of an EDP, there are limitations on the sources and detectors to which it applies. In the entanglement-based protocol, Alice and Bob both measure qubits, in either the $X$ basis or the $Z$ basis --- what we will call {\em standard measurements}. In the corresponding prepare and measure protocol, Alice's source need not emit a qubit, but whatever it emits can be simulated by a standard measurement performed on half of a bipartite state \cite{GottesmanPreskill01}. The state that arrives at Bob's detector also might not be a qubit, but the measurement can be realized as a standard measurement preceded by an operation that ``squashes'' the incoming state to a two-dimensional Hilbert space. 

To be more specific, the source model that we adopt is as follows: Alice's source emits a state in a Hilbert space $A$, where $A$ can be arbitrary, and she launches a state by acting on an auxiliary {\em qubit} $A'$. Alice's basis choice $a\in\{0,1\}$ is determined by flipping a coin. Then a state $\rho_a$ of ${\cal H}_A\otimes {\cal H}_{A'}$, which can depend on the basis $a$, is prepared by Fred. Alice proceeds to perform a standard measurement on her qubit $A'$ in the basis indicated by $a$; that is, a Hadamard transformation is performed on $A'$ if and only if $a=1$, and then Alice measures the qubit in the $Z$ basis. Her measurement determines her key bit: $g=0$ for outcome $+1$, $g=1$ for outcome $-1$.  (Note that, depending on the state $\rho_{a}$, the key bits $g=0,1$ need not be equiprobable.) If Fred's states $\rho_0$ and $\rho_1$ are close to one another, then the states emitted by the source, averaged over the key bit, depend only weakly on the basis.

Actually we can generalize this source model to allow successive emissions to be correlated with one another. Now let $A'$ denote a system of $n$ qubits, $A$ the system in which Alice's $n$ signals reside, and $\rho_a$ a state of 
${\cal H}_A\otimes {\cal H}_{A'}$; the state $\rho_a$ may depend on the $n$-bit string $a$ that specifies Alice's basis choice for each of the $n$ signals. Alice applies a Hadamard transformation to the $i$th qubit if and only if $a_i=1$, then measures the qubits in the $Z$ basis. The measurement outcomes determine her $n$-bit key $g$. Some of the results we report in this paper (Theorem 2, for example) apply to this more general source model. 


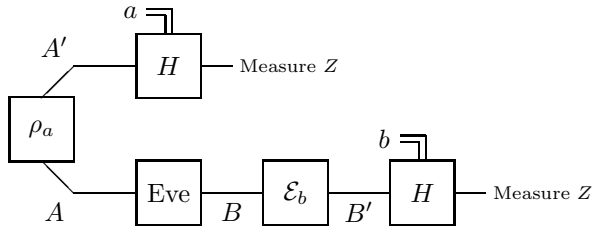
\begin{figure}
\begin{center}
\setlength{\unitlength}{1.2pt}
\begin{picture}(210,73)
\put(50,0){\framebox(20,20){$~$Eve }}
\put(90,0){\framebox(20,20){$~{\cal E}_b$ }}
\put(130,0){\framebox(20,20){$~H$ }}
\put(50,40){\framebox(20,20){$~H$ }}
\put(10,20){\framebox(20,20){$~{\rho}_a$ }}

\put(178,4){\makebox(0,12){\scriptsize Measure $Z$}}
\put(98,44){\makebox(0,12){\scriptsize Measure $Z$}}
\put(128,21){\makebox(0,12){$b$}}
\put(48,61){\makebox(0,12){$a$}}

\put(24,50){\makebox(0,12){$A'$}}
\put(24,-2){\makebox(0,12){$A$}}

\put(80,-2){\makebox(0,12){$B$}}
\put(120,-2){\makebox(0,12){$B'$}}

\put(30,10){\line(1,0){20}}
\put(70,10){\line(1,0){20}}
\put(110,10){\line(1,0){20}}
\put(150,10){\line(1,0){10}}
\put(30,10){\line(-1,1){10}}
\put(20,40){\line(1,1){10}}
\put(30,50){\line(1,0){20}}
\put(70,50){\line(1,0){10}}

\put(139,20){\line(0,1){6}}
\put(141,20){\line(0,1){8}}
\put(139,26){\line(-1,0){6}}
\put(141,28){\line(-1,0){8}}

\put(59,60){\line(0,1){6}}
\put(61,60){\line(0,1){8}}
\put(59,66){\line(-1,0){6}}
\put(61,68){\line(-1,0){8}}

\end{picture}

\end{center}
\caption{Roles of Alice, Bob, Eve, and Fred in the key distribution protocol. Fred prepares an entangled state $\rho_a$ (which may depend slightly on Alice's basis choice $a$) of Alice's $n$-qubit Hilbert space $A'$ and the signal space $A$; then Alice triggers the source by performing a standard measurement on  $A'$. Eve applies an arbitrary basis-independent attack, and then Fred applies a channel ${\cal E}_b$ (which may depend slightly on Bob's basis choice $b$) that ``squashes'' the signals from the Hilbert space $B$ to the $n$-qubit space $B'$. Finally, the qubits in $B'$ are subjected to a standard measurement by Bob. The goal of the protocol is to generate a key that is not known by Eve and Fred, who may communicate freely after Bob measures.}
\label{fig:EveAndFred}
\end{figure}


The $n$ signals emitted by the source are attacked by Eve, who sends to the detector a state that lives in a Hilbert space ${\cal H}_B$. We model the detector as follows: Bob's basis choice $b\in \{0,1\}^{\otimes n}$ for the $n$ signals is determined by flipping $n$ coins. Then Fred applies to the state received by the detector a quantum channel ${\cal E}_b$ that ``squashes'' ${\cal H}_B$ to the $n$-qubit space ${\cal H}_{B'}$; this squash operation may depend on the basis $b$. Bob proceeds to perform standard measurements on the qubits; a Hadamard is performed on the $i$th qubit if and only if $b_i=1$, then Bob measures the qubits in the $Z$ basis. The measurements determine his $n$-bit key $h$ : $h_i=0$ if the outcome of the measurement of the $i$th qubit is $+1$ and $h_i=1$ for outcome $-1$. Since the channel taking ${\cal H}_B$ to ${\cal H}_{B'}$ can act collectively on the incoming signals, our model allows the detector to perform a collective measurement on the $n$ signals it receives. The basis-dependence in the detector's performance is encoded in Fred's channel ${\cal E}_b$.

The prepare-and-measure BB84 protocol, for the source and detector models we have described, is depicted in Fig.~\ref{fig:EveAndFred}. It can be related to a protocol in which entangled pairs of qubits are prepared by Eve (with help from Fred). Half of each pair is delivered to Alice, half to Bob, and they then proceed to perform standard measurements. The security of this latter protocol follows from the security of the corresponding EDP. Therefore, for this model of source and detector, we can use the Shor-Preskill method to analyze the security of BB84.

It may be instructive to contrast our models of the source and detector with those considered in the proofs of Mayers \cite{Mayers96} and Koashi and Preskill \cite{KoashiPreskill02}. Mayers allows the detector to perform an arbitrary two-outcome POVM on each signal it receives, while in our model the POVM must be one that can be realized by a squash followed by a standard measurement. In principle, our model entails no loss of generality, since the Mayers POVM could be followed by the preparation of a qubit in a state chosen so that the standard measurement will reproduce the outcome of the POVM. However, our security proof works only if the channel ${\cal E}_b$ applied by Fred depends sufficiently weakly on the basis $b$, while Mayers requires no such condition. Koashi and Preskill consider a source that can be realized by the preparation of a {\em basis-independent} state of an bipartite system, followed by an arbitrary two-outcome POVM on half of the system, while in our model the POVM must be a standard measurement of a qubit. For the signals emitted by a general Koashi-Preskill source, though it would be possible to launch the same signals by performing a standard measurement on a qubit, this can be done only by choosing bipartite states that depend strongly on the basis, and our security proof works only when the dependence of the states on the basis is sufficiently weak. Therefore, our analysis of security does not apply to the general Mayers detector or the general Koashi-Preskill source. On the other hand, Mayers does not allow Fred to attack the source, Koashi and Preskill do not allow Fred to attack the detector, and the signals emitted by the Koashi-Preskill source reveal no information about the basis used. In contrast, our model allows the performance of both the source and the detector to depend on the basis, and allows the source to leak some information about the basis.

Another noteworthy difference between our model and those of Mayers and Koashi-Preskill is that our model allows Alice's source to emit successive signals that are entangled with one another, and allows the detector to measure the signals collectively. In contrast, Koashi and Preskill assume that the signals emitted by Alice's source are unentangled with one another, and Mayers likewise assumes that the signals are detected by Bob individually rather than collectively. This assumption is used because a crucial step in the Mayers proof is to show that Eve's information about Alice's key would be unchanged if Bob were to flip the basis in which he measures the key bits, but not the basis in which he measures the test bits. A general collective measurement of the signals by Bob would generate correlations between key bit measurements and test bit measurements; therefore, when Bob announces the outcome of his measurements of the test bits he might reveal to Eve some information about his choice of basis for the measurement of the key bits. For this reason, we do not know how to justify the invariance of Eve's information under the basis flip in the case of a collective measurement (though it is not inconceivable that the argument can be extended to cover that case). Similarly, the Koashi--Preskill proof uses the property that Eve's information about Bob's key would be unchanged if Alice were to flip the basis in which she sends the key bits but not the basis in which she sends the test bits, which cannot be justified unless Alice's signals are unentangled.

Our model of the detector can easily be generalized by endowing the detector with imperfect efficiency, so that it sometimes misfires and fails to record an outcome. One simple modification attaches an additional flag bit $d_i$ to each of Bob's qubits. If $d_i=0$, then the $i$th qubit is measured as above, but if $d_i=1$ then the $i$th qubit is discarded and no outcome is recorded.

Detector inefficiencies and other types of losses can be incorporated into the Shor-Preskill security analysis easily enough. Through public discussion, Alice and Bob can eliminate from their sifted key all signals for which Bob failed to record a measurement result. In the entanglement-based protocol, then, we consider an EDP applied to all the pairs from which sifted key bits will be successfully extracted when the measurements are performed. That is, before the EDP is applied we discard all pairs for which Alice and Bob chose different bases or for which the detector misfired, as well as the pairs consumed by the verification test. Security is then proven if we can infer from the test that, with high probability, the remaining pairs have sufficiently low rates of bit errors and phase errors. However, this inference must take into account any basis dependence in the detector efficiency that might contribute to the gap between $\tilde\delta$ and $\tilde\delta_p$, as we will discuss further in Sec.~\ref{sec:pony}. Basis-dependent detector inefficiencies are more problematic for the Mayers argument, since the basis dependence may spoil the invariance of Eve's information about Alice's key when Bob flips his basis for the key bits (but not the test bits).

\section{Real devices}
\label{sec:real}

We are interested in analyzing the security of quantum key distribution with imperfect equipment because we seek assurance that our protocols are secure not just in an ideal world but also in the real world. Therefore, the inherent limitations of our source and detector models should be soberly contemplated.

For example, real sources typically emit not qubits but bosonic modes of the electromagnetic field, and if the likelihood that a mode is multiply occupied is too high, security may be compromised. To evaluate this security threat in our limited framework, we will need to adjust our source model (as we will discuss in Sec.~\ref{sec:tagged}) to incorporate the relevant features, even if not all the detailed physics, of the real source. 

A similar comment applies to detectors. In a typical detector setup for BB84, the incoming photonic mode encounters a polarizing beam splitter that routes the $Z=1$ and $Z=-1$ polarization states (or the $X=\pm 1$ states) to two different photon detectors --- {\em threshold} detectors that do not distinguish one photon from many. If one or the other detector fires, the polarization state is identified. But if more than one photon is present, both detectors might fire, an ambiguous result. If Bob is equipped with such a detector, Eve can trigger the ambiguous result at will by flooding the detector with photons. Even more troubling, Eve can arrange that Bob receive the ambiguous result if he chooses one basis but not the other. For example, Eve can intercept and measure in the $Z$ basis the signal emitted by Alice, and then send on to Bob many $Z$-polarized photons in the state she detects. Then Bob will reproduce Eve's result if he measures in the $Z$ basis, but will obtain the ambiguous result if he measures in the $X$ basis \cite{Norbert98}. Thus, by exploiting the flaw in the design of the detector, Eve can launch a ``Trojan horse'' attack, in effect switching Bob's detector off when it is poised to detect eavesdropping \cite{Lo01}. 
Although our detector model may not fully incorporate all the physics of the polarization beam splitter, we will nonetheless be able to investigate in Sec.~\ref{sec:pony} the power of a Trojan horse attack within an EDP framework.

\section{Choosing the basis quantumly}

For a security analysis that is applicable to BB84 performed with imperfect equipment, we wish to bound the adversary's information in the case of an attack that depends weakly on the basis used to send and detect the signals. For this purpose, we should find a precise formulation of what it means for the basis dependence to be ``weak.'' Therefore, let us focus attention on the coins that Alice and Bob flip to determine their random choices of basis. An attack that depends weakly on the basis is one that depends only slightly on the outcomes of the coin flips.

In the entanglement-based protocol as we have described it up to now, the coin flip is treated classically, and the outcome of the flip determines whether a Hadamard transformation is applied to a qubit before it is measured in the $Z$ basis. Denote by $a_i\in \{0,1\}$ the outcome of the flip of the $i$th coin and by the length $n$ string $a$ the outcome of the flip of $n$ coins. (In the BB84 protocol, Alice and Bob flip separate coins. But for our security analysis we may confine our attention to the sifted key, for which their coin flips agree; therefore in effect there is only one basis choice $a_i$ for each signal.) Denote by $H(a)$ the operation which applies a Hadamard to the $i$th qubit if $a_i=1$ and the identity to the $i$th qubit if $a_i=0$. Then in the setting where Eve knows nothing about the basis choice, the effect of the randomly applied Hadamards by Alice and Bob (after the attack by Eve) is to transform the state of the $n$ pairs according to
\begin{equation}
\label{Hadamard_on_a}
\rho\to \rho'={1\over 2^n}\sum_{a=0}^{2^n-1} \left(H(a)\otimes H(a)\right) \rho \left(H(a)\otimes H(a)\right)~.
\end{equation}
Then since $H(b)H(a)=H(a\oplus b)$, $\rho'$ has the property of Hadamard invariance: for any bit string $b$,
\begin{equation}
 \left(H(b)\otimes H(b)\right) \rho' \left(H(b)\otimes H(b)\right)=\rho' ~.
\end{equation}
In the Shor-Preskill argument, this symmetry of $\rho'$ is used to infer that the bit error rate and phase error rate of the key generating pairs are, with high probability, nearly the same.

In order to analyze (weakly) basis-dependent attacks, it is convenient to treat the coin flip quantumly rather than classically --- we can imagine that each coin is in a coherent superposition of heads and tails, and that the Hadamard transform is conditioned on the state of the coins. In the ideal protocol, the $n$ coins are prepared in the state 
\begin{equation}
\left({1\over \sqrt{2}}(|0\rangle+|1\rangle)\right)^{\otimes n}= {1\over \sqrt{2^n}}\sum_{a=0}^{2^n-1}|a\rangle~,
\end{equation}
and the Hadamard is applied to the $i$th pair if $a_i=1$ --- therefore, if $|\Psi\rangle$ is the state of the pairs, then the effect of the random basis choice can be expressed as
\begin{equation}
|\Psi\rangle \otimes |a\rangle \to  \left(H(a)\otimes H(a)\right)|\Psi\rangle \otimes |a\rangle~.
\end{equation}
When we trace over the state of the coin, the effect on the quantum state of the pairs is just as in eq.~(\ref{Hadamard_on_a}).

\begin{figure}
\begin{center}
\begin{picture}(220,60)
\put(60,0){\framebox(40,40){$~$Fred }}
\put(120,0){\framebox(20,20){$~H$ }}

\put(40,10){\line(1,0){20}}
\put(40,30){\line(1,0){20}}
\put(100,10){\line(1,0){20}}
\put(100,30){\line(1,0){50}}
\put(140,10){\line(1,0){10}}
\put(130,30){\line(0,-1){10}}

\put(120,52){\vector(-1,-2){10}}
\put(120,52){\line(1,0){10}}

\put(130,30){\circle*{4}}

\put(176,4){\makebox(0,12){Measure $Z$}}
\put(176,24){\makebox(0,12){Measure $Z$}}
\put(150,46){\makebox(0,12){\scriptsize Measure $X$}}
\put(24,24){\makebox(0,12){coin}}
\put(24,4){\makebox(0,12){signal}}

\end{picture}

\end{center}
\caption{The quantum coin. The basis choice for the detector (and the source) is determined by measuring a qubit (the coin) in the $Z$ basis. To generate each bit of sifted key, a conditional Hadamard transformation, controlled by the coin, is applied to the signal qubit, and then the signal qubit is measured in the $Z$ basis. Fred's basis-dependent attack on the signal can be described as a joint attack on the coin and the signal. To quantify how the coin is disturbed by Fred's attack, we consider measuring the coin in the $X$ basis after the attack and before the conditional Hadamard.}
\label{fig:FredandCoin}
\end{figure}
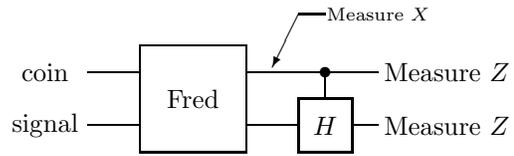

Now, in this formulation, it is easy to describe the distinction between Eve's basis-independent attack and Fred's basis-dependent attack. Eve interacts only with the pairs, but Fred is permitted to tamper with both the pairs {\em and} the coins, as in Fig.~\ref{fig:FredandCoin}. In the actual protocol, the coin is classical, but it will not make Fred any less powerful if we allow him to attack a quantum coin instead. (When we say that the coin is ``classical,'' we mean Fred's attack is a quantum operation applied to the pairs that is conditioned on the state of the coin in a preferred basis. We will prove security for general attacks by Fred with weak dependence on the state of the coin, so our results will apply in particular to the case of a classical coin.) Furthermore, it is easy to state precisely what it means for the attack to depend only weakly on the basis: the basis dependence is weak if Fred's attack disturbs the coin only slightly. This notion of weak basis dependence applies even if we allow Fred to attack the signals twice, at the source (before Eve's attack) and at the detector (after Eve's attack).  Actually, once we introduce the quantum coin in this way, it is not so important to keep Fred in the picture at all --- we can go back to the usual picture in which there is only one adversary, but limit Eve's attack on the coin. 

\medskip
\noindent{\bf Definition. $\Delta$-balanced attack}. {\em Suppose that after $n$ pairs and the $n$ corresponding coins are attacked by the adversary (but before the final Hadamard transformations, conditioned on the coins, that precede the measurement of the pairs in the $Z$ basis), the $n$ coins are all measured in the $X$ basis. The attack is $\Delta$-balanced if, with high probability, the number of coins for which the measurement outcome is $X=-1$ is less than $n\Delta$.}
\medskip

\noindent 
If $\Delta\approx 0$, the attack is {\em balanced} --- that is, basis-independent.
We will prove in Sec.~\ref{sec:generic-proof} that if the attack is $\Delta$-balanced and $\Delta$ is sufficiently small, then secure quantum key distribution is possible, and we will obtain a lower bound on the achievable key generation rate. Later we will discuss some more specific examples of $\Delta$-balanced attacks, and for some of those attacks we will obtain stronger lower bounds on the rate.

We emphasize again that for a security analysis it suffices to imagine that Alice and Bob share a {\em single} quantum coin that determines the choice of basis for each signal. 
Of course, in the BB84 protocol, Alice and Bob use separate classical coins to determine whether to use the $Z$ basis or the $X$ basis. But the quantum coin is not intended to provide an accurate portrayal of the actual protocol; it is a mathematical device for analyzing the impact of the basis dependence of the attack. For the purpose of this analysis, we  replace the two classical coins by a single quantum coin only after discarding the cases in which Alice's classical coin flip  and Bob's classical coin flip yield different outcomes. 

We have seen that for the analysis of the BB84 protocol, it is convenient to imagine that Alice delays the measurement that launches her signals until after Eve's attack. That way, we can relate the classical privacy amplification in BB84 to an EDP, and so establish security. Here we are taking this idea a step further. It is convenient to imagine that the measurement of the coin that determines the basis is also delayed until after the attack by Eve and Fred. That way, we can infer a bound on the asymmetry between the bit error rate and the phase error rate for pairs subjected to the EDP, and so establish that the EDP will be effective.

\section{Security proof for small basis-dependent flaws}
\label{sec:generic-proof}
To analyze security, we'd like to relate the asymmetry of the coin (as parametrized by $\Delta$) to the gap between the bit error rate $\tilde\delta$ and the phase error rate $\tilde\delta_p$ when the pairs are measured in the Bell basis. First we need to write down a convenient expression for this gap. 

In the entanglement-based protocol, the random variables $n\tilde\delta$ and $n\tilde\delta_p$ are defined as the number of bit errors and phase errors that would be found if the $n$ key generating pairs were all measured in the Bell basis. For a particular pair, consider the observable 
\begin{equation}
{1\over 2}\left(Z\otimes Z - X\otimes X\right)= |\phi^-\rangle\langle \phi^-| - |\psi^+\rangle\langle \psi^+|~,
\end{equation} 
where $|\phi^-\rangle$ and $|\psi^+\rangle$ are the Bell states
\begin{eqnarray}
|\phi^-\rangle&=& {1\over\sqrt{2}}\left(|00\rangle - |11\rangle\right)~,\nonumber\\
|\psi^+\rangle&=& {1\over\sqrt{2}}\left(|01\rangle + |10\rangle\right)~.
\end{eqnarray}
This observable has eigenvalues $\{+1,0,-1\}$. If the pair is to be measured in the $Z$ basis, we say that there is a bit error if $Z\otimes Z=-1$ and that there is a phase error if $X\otimes X=-1$. Therefore, the eigenvalue of ${1\over 2}(Z\otimes Z- X\otimes X)$ is $-1$ if there is a bit error but no phase error, $+1$ if there is a phase error but no bit error, and $0$ if either there are no errors or both a bit error and a phase error. Similarly, if the pair is to be measured in the $X$ basis, we say that there is a phase error if $Z\otimes Z=-1$ and that there is a bit error if $X\otimes X=-1$. Therefore, the eigenvalue of ${1\over 2}(Z\otimes Z- X\otimes X)$ is $+1$ if there is a bit error but no phase error, $-1$ if there is a phase error but no bit error, and $0$ if either there are no errors or both a bit error and a phase error. 

Suppose that the basis choice is decided by flipping a coin, where the pairs are to be measured in the $Z$ basis if the outcome of the coin flip is $|0\rangle$ ($Z_{\rm coin}=1$), and the pairs are to be measured in the $X$ basis if the outcome of the coin flip is $|1\rangle$ ($Z_{\rm coin}=-1$). Then the observable
\begin{equation}
{1\over 2}\left(Z\otimes Z - X\otimes X\right)_{\rm pair}\otimes Z_{\rm coin}
\end{equation}
has the eigenvalue $-1$ if the pair has a bit error but no phase error, the eigenvalue $+1$ if the pair has a phase error but no bit error, and the eigenvalue $0$ otherwise. We see, then, that for the $n$ key generating pairs, the gap between the number of phase errors and the number of bit errors can be expressed as
\begin{eqnarray}
\label{n_gap}
&&n_{\rm gap}\equiv n\left(\tilde\delta_p -\tilde\delta\right)\nonumber\\
&&=\sum_{i=1}^n {1\over 2}\left(Z\otimes Z - X\otimes X\right)_{{\rm pair}, i}\otimes Z_{{\rm coin},i}~.
\end{eqnarray}
Eq.~(\ref{n_gap}) means that $n_{\rm gap}$ is a random variable whose probability distribution is the distribution of outcomes if the observable on the right-hand side of eq.~(\ref{n_gap}) is measured. We might imagine that $Z\otimes Z$ and $X\otimes X$ are measured for every pair (this is a complete Bell measurement) and that $Z_{\rm coin}$ is measured for every coin; then $n_{\rm gap}$ is found by summing up all the results of these measurements. But since the Bell measurements and the coin measurements all commute with our expression for $n_{\rm gap}$ in eq.~(\ref{n_gap}), we could just as well imagine that $n_{\rm gap}$ is measured first, and that the other measurements are completed later --- the probability distribution for $n_{\rm gap}$ will be the same either way. In any case, our expression for $n_{\rm gap}$ is valid even if there are strong correlations among the pairs and the coins.

If we imagine that all of the coins are measured in the $X$ basis (as in the definition of a $\Delta$-balanced attack), then the random variable that represents the number of coins for which the outcome is $X=-1$ can be expressed as
\begin{equation}
\label{N_minus}
n_{\rm coin}^{(X)}=\sum_{i=1}^n \left(I\otimes I\right)_{{\rm pair}, i}\otimes {1\over 2}(I-X)_{{\rm coin},i}~.
\end{equation}
We wish to obtain a bound on $n_{\rm gap}$ that will hold with high probability for any possible state of the pairs and the coins such that $n_{\rm coin}^{(X)}$ is less than $n\Delta$ with high probability. 
It is convenient to express the gap as a sum of two terms, $n_{\rm gap}=n_{\rm gap}^{(Z)}+n_{\rm gap}^{(X)}$, and to bound each term separately. First, consider
\begin{equation}
n_{\rm gap}^{(Z)}={1\over 2}\sum_{i=1}^n \left(Z\otimes Z\right)_{{\rm pair}, i}\otimes Z_{{\rm coin},i}~.
\end{equation}
For each value of $i$, imagine that we perform two successive controlled-NOT gates, one with Alice's qubit as the control and the coin as the target, and the other with Bob's qubit as the control and the coin as the target. Acting by conjugation, the effect of these gates is
\begin{eqnarray}
&&(Z\otimes Z)_{\rm pair} \otimes Z_{\rm coin} \to (I\otimes I)_{\rm pair} \otimes Z_{\rm coin}~, \nonumber\\
&&(I\otimes I)_{\rm pair} \otimes X_{\rm coin} \to (I\otimes I)_{\rm pair} \otimes X_{\rm coin}~.
\end{eqnarray}
Therefore, this change of basis has no effect on the statistics of the observable $n_{\rm coin}^{(X)}$, while transforming the observable 
$n_{\rm gap}^{(Z)}$ according to
\begin{equation}
\label{gap_trans}
 n_{\rm gap}^{(Z)}\to {n\over 2} - n_{\rm coin}^{(Z)}~,
\end{equation}
where 
\begin{equation}
n_{\rm coin}^{(Z)}=\sum_{i=1}^n \left(I\otimes I\right)_{{\rm pair}, i}\otimes{1\over 2}(I-Z)_{{\rm coin},i}~
\end{equation}
(the number of coins for which $Z=-1$, if all $n$ are measured in the $Z$ basis).

We are interested in analyzing how the statistics of $n_{\rm gap}^{(Z)}$ is related to the statistics of $n_{\rm coin}^{(X)}$. Let $\rho$ denote the state of the $n$ coins and the $n$ pairs, and suppose that the controlled-NOT gates described above transform this state to a new state $\rho'$. We see that the statistics of $n_{\rm coin}^{(X)}$ and $n_{\rm gap}^{(Z)}$ in the state $\rho$ is identical to the statistics of $n_{\rm coin}^{(X)}$ and ${n\over 2} - n_{\rm coin}^{(Z)}$ in the state $\rho'$. Therefore, to derive a relation between $n_{\rm coin}^{(X)}$ and $n_{\rm gap}^{(Z)}$ that holds with high probability for an arbitrary state $\rho$, it suffices to analyze how $n_{\rm coin}^{(Z)}$ and $n_{\rm coin}^{(X)}$ are related for an arbitrary state $\rho'$. For this we appeal to the following lemma, which asserts that if $n_{\rm coin}^{(X)}$ is small, then $n_{\rm coin}^{(Z)}$ is close to $n/2$:

\medskip
\noindent{\bf Lemma 1} {\em For a quantum state of $n$ coins, suppose that with high probability $n_{\rm coin}^{(X)}<n\Delta$. Then, for any positive $\varepsilon$, with high probability 
\begin{equation}
\label{Z_bound}
|n/2 - n_{\rm coin}^{(Z)}| < n(f(\Delta)+\varepsilon)~,
\end{equation}
where $\Delta$ and $f(\Delta)$ are related by
\begin{equation}
\label{f_define}
H_2(1/2 -f(\Delta)) + H_2(\Delta) = 1~.
\end{equation} 
}
\medskip

\noindent {\bf Proof}: 
The proof for the case of a pure quantum state $|\psi\rangle$ of $n$ coins, where $n^{(X)}<n\Delta$ with probability 1, is in Appendix A of \cite{GottesmanChuang}. But if instead $n^{(X)}<n\Delta$ with high probability, we can write $|\psi\rangle= |\psi\rangle_{\rm good} + |\psi\rangle_{\rm bad}$, where the (unnormalized) state $|\psi\rangle_{\rm good}$ has $n^{(X)}<n\Delta$ with probability 1, and $\parallel\psi_{\rm bad}\parallel$ is exponentially small. Hence $|\psi\rangle_{\rm good}$, and therefore also $|\psi\rangle$, has the property eq.~(\ref{Z_bound}) with high probability. Therefore, Lemma 1 holds for pure states. Now, a mixed state can be realized as an ensemble of pure states. By the hypothesis of Lemma 1, all of the pure states in this ensemble, except for those occuring with exponentially small probability, satisfy $n^{(X)}<n\Delta$ with high probability, and therefore also satisfy eq.~(\ref{Z_bound}) with high probability. This proves Lemma 1.

We note that by expanding $H_2(1/2 -f)$ as a power series in $f$, and using the convexity of $H_2$, we can derive from eq.~(\ref{f_define}) a useful inequality satisfied by $f(\Delta)$:
\begin{equation}
\big(f(\Delta)\big)^2 \le {1\over 2}(\ln 2) H_2(\Delta)~.
\end{equation}
Expanding this expression for small $\Delta$ and using convexity again, we obtain
\begin{equation}
f(\Delta) \le \sqrt{{\Delta\over 2}\ln\left({e\over \Delta}\right)}~.
\end{equation}

From Lemma 1 and eq.~(\ref{gap_trans}), we infer that, for a $\Delta$-balanced attack, $|n_{\rm gap}^{(Z)}| < n(f(\Delta)+\varepsilon)$ with high probability. A similar argument shows that also $|n_{\rm gap}^{(X)}| < n(f(\Delta) + \varepsilon)$ with high probability, where 
\begin{equation}
n_{\rm gap}^{(X)}=-{1\over 2}\sum_{i=1}^n \left(X\otimes X\right)_{{\rm pairs}, i}\otimes Z_{{\rm coin},i}~.
\end{equation}
(For this argument, we apply Hadamard transformations to all pairs before applying the CNOT gates.) Since $n_{\rm gap} = n_{\rm gap}^{(Z)}+n_{\rm gap}^{(X)}$, we have proved:

\medskip
\noindent{\bf Lemma 2} {\em For a $\Delta$-balanced attack on $n$ pairs and $n$ coins, the state of the pairs has the property
\begin{equation}
|\tilde\delta_p-\tilde\delta| < 2(f(\Delta)+\varepsilon)~
\end{equation}
with high probability, for any positive $\varepsilon$.}
\medskip

With Lemma 2 in hand, we can now complete the proof of security following the steps outlined in Sec.~\ref{sec:shor-preskill}. If the error rate found in the test is $\delta$, then the number of bit errors in the key-generating pairs is less than $n(\delta+\varepsilon)$ with high probability (assuming that Eve's attack passes the test with a probability that is not exponentially small). For a $\Delta$-balanced attack, we infer that the number of phase errors in the key-generating pairs is less than $n(\delta +2f(\Delta) +\varepsilon)$ with high probability.  By introducing a random permutation (not known by Eve or Fred) we can ensure that the errors are randomly distributed among the pairs. Therefore, for a suitable CSS code, high fidelity pairs (and hence secure key) can be extracted at a rate $1- H_2(\delta+\varepsilon) - H_2(\delta + 2f(\Delta) +\varepsilon)$, for any positive $\varepsilon$. We have proved:

\medskip
\noindent{\bf Theorem 2. Security of BB84 against weakly basis-dependent attacks}. {\em The BB84 protocol is secure if Eve and Fred launch a $\Delta$-balanced attack. Secure final key can be extracted from sifted key at the asymptotic rate
\begin{equation}
\label{general_key_rate}
R= {\rm Max}\big(1- H_2(\delta) - H_2(\delta + 2f(\Delta)),0\big)
\end{equation}	  		                     
where $\delta$  is the bit error rate found in the verification test and $f(\Delta)$ is defined as in eq.~(\ref{f_define}).} ({\em We assume $\delta + 2f(\Delta)<1/2$.})
\medskip

\noindent We note that the key generation rate found in Theorem 2 is nonzero only for $2f(\Delta) < 1/2$, or $\Delta < .0289$.

Theorem 2 is our central result concerning security for equipment with generic flaws. In the remainder of this paper, we will analyze some specific examples. As we will see, for some special cases we can establish a key generation rate exceeding the rate eq.~(\ref{general_key_rate}) found for the general case.

\section{Individual source flaws and a perfect detector}
\label{sec:individual_source}
As our first application of Theorem 2, we consider the case where the detector is perfect, but the source is subject to individual flaws that may leak some information to Eve about Alice's basis choice. We will prove security by showing that the attack is $\Delta$-balanced.

Suppose that Alice's source emits one of four possible states of a single qubit. In the ideal protocol, these states are the four BB84 states, chosen equiprobably. Suppose, though, that the source is imperfect, so that the four states differ from the corresponding BB84 states, but only slightly.

Let $a\in\{0,1\}$ denote Alice's declared basis choice (ideally, the $Z$ basis for $a=0$ and the $X$ basis for $a=1$) and let $g\in\{0,1\}$ denote Alice's key bit. Suppose that $a$ and $g$ are chosen with the joint probability $p_{a,g}$, and that once the values of $a$ and $g$ are chosen, Alice's source emits a state $\rho_{a,g}$. 
The Koashi-Preskill analysis applies if $p_{0,0}\rho_{0,0}+p_{0,1}\rho_{0,1}=p_{1,0}\rho_{1,0}+p_{1,1}\rho_{1,1}$, the case in which the source does not reveal any information about $a$. We will say that the source is {\em oblivious} when it has this property. Now we are interested in the case were the source is nonoblivious --- it leaks a small amount of information about the basis choice. 

We can characterize the flawed source by imagining that Alice prepares her states by performing an ideal measurement on half of an entangled pair. The state of the pair (prior to Alice's measurement) is $\rho_0$ for $a=0$ and $\rho_1$ for $a=1$. The basis-dependence of the source is weak in the sense that the states $\rho_0$ and $\rho_1$ differ only slightly --- their fidelity is close to one:
\begin{equation}
\label{source_fidelity}
\sqrt{F(\rho_0,\rho_1)}\equiv ~ \parallel \sqrt{\rho_0}\sqrt{\rho_1}\parallel_{\rm tr} ~>  1-2\varepsilon_s~.
\end{equation}
If $n$ signals are sent, the state that Fred prepares is a product state: $\bigotimes_{i=1}^n \rho^{(i)}_{a_i}$, where $a_i$ denotes the basis choice for the $i$th signal, and $\sqrt{F(\rho^{(i)}_0,\rho^{(i)}_1)}> 1-2\varepsilon_s$ for each $i$. Thus we say that Fred's attack on the source is {\em individual}, and that the basis-dependence is characterized by $\varepsilon_s$. We will suppose for now that any flaws in the detector are basis independent, so that Fred attacks only the source.

The states $\rho_0$ and $\rho_1$ may be mixed in general, but they can be ``purified'' by introducing a suitable ``environment'' $E$; that is, there are pure states $|\Psi_0\rangle$ and $|\Psi_1\rangle$ such that
\begin{equation}
{\rm tr}_E \big(|\Psi_0\rangle\langle \Psi_0|\big)=\rho_0~,\quad {\rm tr}_E \big(|\Psi_1\rangle\langle \Psi_1|\big)=\rho_1~.
\end{equation}
Furthermore, it follows from eq.~(\ref{source_fidelity}) that the purifications can be chosen to have a large overlap \cite{uhlmann76,jozsa94}:
\begin{equation}
{\rm Re} ~\langle \Psi_1|\Psi_0\rangle > 1-2\varepsilon_s~.
\end{equation}

Now suppose that, as in Sec.~VI,  we imagine that the basis choice is determined by a ``quantum coin.'' Then, the state of the coin, the pair, and the environment can be described as a pure state
\begin{equation}
\label{coin_and_pair}
{1\over\sqrt{2}}\big(|\Psi_0\rangle\otimes |0\rangle + |\Psi_1\rangle \otimes |1\rangle\big)~.
\end{equation}
If the state of the pair used by Alice to prepare her signal depends on the choice of basis, then the coin will be entangled with the pair and environment, and the strength of this entanglement will depend on how much $|\Psi_0\rangle$ and $|\Psi_1\rangle$ differ. Of course, the quantum coin is merely a mathematical fiction that we invoke for the purpose of analyzing the basis dependence of the pairs that are used to generate the key in the entanglement-based key distribution protocol. Furthermore, the state of the pairs does not depend on how we choose the purifications of $\rho_0$ and $\rho_1$. But the state of the coins {\em does} depend on this choice, and we may exploit our freedom in choosing the purifications to obtain the strongest possible bound on the basis dependence of the pairs.

Since we are assuming that any flaws in the detector are basis independent, these may be absorbed into Eve's basis-independent attack. Then since Eve's attack has no effect on the coins, the state of any coin can be completely characterized by tracing out the pair and environment from eq.~(\ref{coin_and_pair}). If the state of the coin is now measured in the $X$ basis, the outcome $X=-1$ occurs with probability
\begin{eqnarray}
\label{x-1_prob}
p&=&{1\over 4}\parallel |\Psi_{0}\rangle-|\Psi_{1}\rangle\parallel ^2\nonumber\\
&=& {1\over 2} \Big(1- {\rm Re}~ \langle\Psi_{1}|\Psi_{0}\rangle\Big) <\varepsilon_s~.
\end{eqnarray}
Because the attack is individual, the coins are independent and this bound on $p$ applies to each one of the $n$ coins; therefore we conclude that the attack is $(\varepsilon_s +\varepsilon)$-balanced, for any positive $\varepsilon$. Hence from Theorem 2 we obtain

\medskip
\noindent{\bf Theorem 3. Security of BB84 for a source with individual weakly basis-dependent flaws}. {\em Suppose that the flaws in the detector are basis-independent, and that the flaws in the source are individual. The $i$th signal sent by Alice is prepared by performing a standard qubit measurement on half of an entangled state --- this state is $\rho_0^{(i)}$ when the $Z$ basis is declared and $\rho_1^{(i)}$ when the $X$ basis is declared, where $\sqrt{F(\rho_0^{(i)},\rho_1^{(i)})} > 1-2\varepsilon_s$ for all $i$. Then the BB84 protocol is secure, and  secure final key can be extracted from sifted key at the asymptotic rate
\begin{equation}
R={\rm Max}\left(1- H_2(\delta)-H_2(\delta + 2f(\varepsilon_s),0\right)
\end{equation}	  		                     
where $\delta$  is the bit error rate found in the verification test  and $f(\varepsilon_s)$ is defined as in eq.~(\ref{f_define}).} ({\em We assume $\delta + 2f(\varepsilon_s)<1/2$.})
\medskip

Note that in the formulation of Theorem 3 we have assumed that all signals are detected --- we have not considered the effects of loss in the channel or imperfect detector efficiency. In principle, Eve can amplify the basis-dependence of Fred's attack by eliminating some of the signals. In the worst case, the coin is an $X=1$ eigenstate for each of the signals that Eve removes. Then, if a fraction $f$ of all the signals are lost, $\Delta$ is enhanced according to
\begin{equation}
\Delta\to \Delta' \le  \Delta/(1-f)~.
\end{equation}
The effects of loss will be discussed further in Sec.~\ref{sec:tagged} and Sec.~\ref{sec:pony}.

\section{Imperfect oblivious source and imperfect detector}
\label{sec:individual_pair}

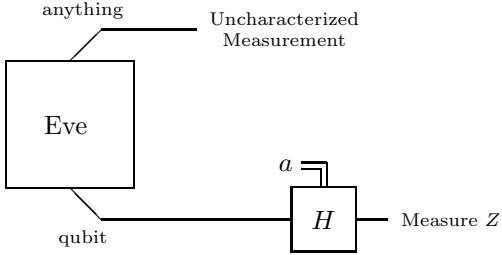
\begin{figure}
\begin{center}
\setlength{\unitlength}{1.2pt}
\begin{picture}(170,78)

\put(10,20){\framebox(40,40){Eve }}

\put(40,10){\line(1,0){60}}
\put(40,10){\line(-1,1){10}}

\put(100,0){\framebox(20,20){$~H$ }}
\put(98,21){\makebox(0,12){$a$}}
\put(109,20){\line(0,1){6}}
\put(111,20){\line(0,1){8}}
\put(109,26){\line(-1,0){6}}
\put(111,28){\line(-1,0){8}}
\put(120,10){\line(1,0){10}}

\put(150,4){\makebox(0,12){\scriptsize Measure $Z$}}
\put(96,64){\makebox(0,12){ ${\rm Uncharacterized}\atop \rm Measurement $}}

\put(34,70){\makebox(0,12){\scriptsize anything}}
\put(34,-2){\makebox(0,12){\scriptsize qubit}}

\put(30,60){\line(1,1){10}}
\put(40,70){\line(1,0){30}}

\end{picture}
\end{center}
\caption{An uncharacterized oblivious source and a perfect detector. Eve prepares an entangled state of $n$ signals and $n$ qubits. Alice prepares an uncharacterized measurement on the $n$ signals and Bob performs standard measurements on the $n$ qubits. Interchanging the roles of Alice and Bob, we obtain the case of a perfect source and an uncharacterized detector. }
\label{fig:Eve'sPairs}
\end{figure}

We recall that Koashi and Preskill \cite{KoashiPreskill02} proved the security of BB84 in the case where the detector is perfect and the signals emitted by the source, when averaged over the key bits, are basis independent (an {\em oblivious} source). The situation they considered can be depicted as in Fig.~\ref{fig:Eve'sPairs}. In effect, Eve prepares an entangled state of $n$ qubits, which are delivered to Bob, and $n$ general signals, which are delivered to Alice. To generate the sifted key, Alice performs an uncharacterized measurement on each of her $n$ signals, and Bob performs a standard measurement on each of his $n$ qubits. By simply reversing the roles of Alice and Bob, we obtain the situation considered by Mayers, in which the source is perfect and the detector is uncharacterized \cite{Mayers96}. 

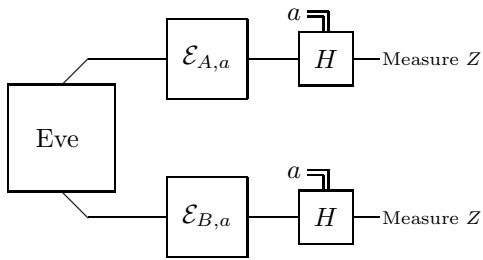
\begin{figure}
\begin{center}
\begin{picture}(190,88)

\put(10,20){\framebox(40,40){Eve }}

\put(40,10){\line(1,0){30}}
\put(40,10){\line(-1,1){10}}

\put(120,0){\framebox(20,20){$~H$ }}
\put(118,21){\makebox(0,12){$a$}}
\put(129,20){\line(0,1){6}}
\put(131,20){\line(0,1){8}}
\put(129,26){\line(-1,0){6}}
\put(131,28){\line(-1,0){8}}
\put(140,10){\line(1,0){10}}
\put(170,4){\makebox(0,12){\scriptsize Measure $Z$}}

\put(120,60){\framebox(20,20){$~H$ }}
\put(118,81){\makebox(0,12){$a$}}
\put(129,80){\line(0,1){6}}
\put(131,80){\line(0,1){8}}
\put(129,86){\line(-1,0){6}}
\put(131,88){\line(-1,0){8}}
\put(140,70){\line(1,0){10}}
\put(170,64){\makebox(0,12){\scriptsize Measure $Z$}}

\put(30,60){\line(1,1){10}}
\put(40,70){\line(1,0){30}}

\put(70,55){\framebox(30,30){$~{\cal E}_{A,a}$ }}
\put(70,-5){\framebox(30,30){$~{\cal E}_{B,a}$ }}

\put(100,70){\line(1,0){20}}
\put(100,10){\line(1,0){20}}

\end{picture}
\end{center}
\caption{An oblivious source and an imperfect detector. }
\label{fig:FredAttacksTwice}
\end{figure}

\begin{figure}
\begin{center}
\begin{picture}(190,78)

\put(10,20){\framebox(40,40){Eve }}

\put(40,10){\line(1,0){30}}
\put(40,10){\line(-1,1){10}}

\put(120,0){\framebox(20,20){$~H$ }}
\put(118,21){\makebox(0,12){$a$}}
\put(129,20){\line(0,1){6}}
\put(131,20){\line(0,1){8}}
\put(129,26){\line(-1,0){6}}
\put(131,28){\line(-1,0){8}}
\put(140,10){\line(1,0){10}}
\put(170,4){\makebox(0,12){\scriptsize Measure $Z$}}

\put(120,60){\framebox(20,20){$~H$ }}
\put(118,81){\makebox(0,12){$a$}}
\put(129,80){\line(0,1){6}}
\put(131,80){\line(0,1){8}}
\put(129,86){\line(-1,0){6}}
\put(131,88){\line(-1,0){8}}
\put(140,70){\line(1,0){10}}
\put(170,64){\makebox(0,12){\scriptsize Measure $Z$}}

\put(30,60){\line(1,1){10}}
\put(40,70){\line(1,0){30}}

\put(70,-5){\framebox(30,95){$~{\cal E}_{a}$ }}

\put(100,70){\line(1,0){20}}
\put(100,10){\line(1,0){20}}

\end{picture}
\end{center}
\caption{Stronger version of the attack in Fig.~\ref{fig:FredAttacksTwice}. }
\label{fig:FredAttacksPairs}
\end{figure}
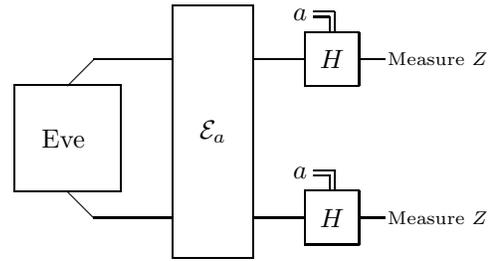

We will now consider a special case of the Koashi-Preskill source: the source is oblivious, but we further assume that the source can be realized by the preparation of a basis-independent entangled state of the signal space and a {\em qubit}, followed by a basis-dependent channel applied to the qubit, and finally a standard measurement of the qubit. However, we will go beyond Koashi and Preskill by allowing the detector to have basis-dependent flaws, as shown in Fig.~\ref{fig:FredAttacksTwice}.  Actually, it will be no harder to analyze the more general case shown in Fig.~\ref{fig:FredAttacksPairs}: Eve prepares an arbitrary state of $n$ entangled signals, which is mapped by Fred to a state of $n$ pairs of qubits; then the pairs are distributed to Alice and Bob, who perform standard measurements. An important feature of this setting is that, although Fred's channel can depend on the basis in which Alice and Bob measure, there is no way for Fred to convey any information about the basis to Eve. In this sense the source is oblivious.

We will further assume that the channel applied by Fred is a product of $n$ individual channels, and that each of these $n$ channels depends only weakly on the basis. For analyzing the impact on the quantum coin, it will be convenient to characterize the basis dependence of Fred's attack as follows: A channel ${\cal E}$ that takes Eve's arbitrary pair to a pair of qubits can be realized by its {\em dilation}, an isometric embedding $U$ of Eve's space into the space of the qubit pair and a suitable ancilla. Thus Fred's basis-dependent individual attack can be expressed as the tensor product $U_a\equiv \bigotimes_{i=1}^n U^{(i)}_{a_i}$, where $i$ labels the pairs, and $a_i$ denotes the basis choice for the $i$th signal. Furthermore the attack depends only weakly on the basis, in the sense that ${1\over 4}\|U^{(i)}_0- U^{(i)}_1\|_{\rm sup}^2< \varepsilon$ for each $i$. 

Using this characterization, we can analyze how Fred's attack affects the coins that determine the basis. 
The basis choice is determined by $n$ quantum coins, each a qubit initially prepared in the $X=1$ eigenstate, and suppose that the initial state of the $n$ pairs and their environment (before Fred's attack) is the pure state $|\varphi\rangle$. Then after Fred's attack, the state of the coins, the pairs, and the environment can be written as
\begin{equation}
{1\over 2^{n/2}}\sum_a U_a|\varphi\rangle\otimes |Z;a\rangle~,
\end{equation}
where $a$ is the $n$ bit string indicating the basis choice, and the states $\{|Z;a\rangle\}$ are the basis states of the coins in the $Z$ basis. 
After Fred's attack, suppose that all of the $n$ coins are measured in the $X$ basis. Let $x$ be an $n$-bit string, and let $|X;x\rangle$ denote a product of $n$ $X$ eigenstates, such that $X_i=1$ for $x_i=0$ and $X_i=-1$ for $x_i=1$. Then the probability that the measurement of the coins yields the outcome $x$ is 
\begin{eqnarray}
P(x)&=&\left\|{1\over 2^{n/2}}\sum_a U_a|\varphi\rangle\otimes \langle X;x|Z;a\rangle\right\|^2~,\nonumber\\
&=&\left\|{1\over 2^{n}}\left(\sum_a (-1)^{a.x}~U_a\right)|\varphi\rangle \right\|^2\nonumber\\
&\le& \left\|{1\over 2^{n}}\left(\sum_a (-1)^{a.x}~U_a\right)\right\|_{\rm sup}^2~.
\end{eqnarray}
The sum in this expression can be factorized:
\begin{equation}
{1\over 2^{n}}\sum_a (-1)^{a.x}~U_a = \bigotimes_{i=1}^n {1\over 2}\left(U_{0}^{(i)} + (-1)^{x_i} U_{1}^{(i)}\right)~.
\end{equation}
Furthermore, the sup norm of a tensor product is a product of sup norms. Since for each $i$
\begin{eqnarray}
\left\| {1\over 2} \left(U_0^{(i)}-U_1^{(i)}\right)\right\|_{\rm sup}^2 < \varepsilon~,\nonumber\\
\left\| {1\over 2} \left(U_0^{(i)}+U_1^{(i)}\right)\right\|_{\rm sup}^2 \le 1~,
\end{eqnarray}
we have
\begin{equation}
\left\|{1\over 2^{n}}\left(\sum_a (-1)^{a.x}~U_a\right)\right\|_{\rm sup}^2 < (\varepsilon)^{|x|}~,
\end{equation}
where $|x|$ denotes the Hamming weight of $x$, and therefore
\begin{equation}
P(x) < (\varepsilon)^{|x|}~.
\end{equation}
From this bound on $P(x)$, it is elementary to show that the probability that $|x| > (e\varepsilon + \varepsilon')n$ is exponentially small for any positive $\varepsilon'$ and $e=2.71828\cdots$. Therefore we have

\medskip
\noindent{\bf Lemma 3.  A weakly basis-dependent individual oblivious attack by Fred is $\Delta$-balanced}. {\em Consider an individual attack by Fred, in which Fred applies $U_0^{(i)}$ to the $i$th pair if the basis choice is $a_i=0$ and applies $U_1^{(i)}$ if the basis choice is $a_i=1$, where   ${1\over 4}\|U^{(i)}_0- U^{(i)}_1\|_{\rm sup}^2< \varepsilon$ for each $i$. This attack is $\Delta-balanced$ for any $\Delta > e\varepsilon$.
}\medskip

And from Theorem 2 we obtain:

\medskip
\noindent{\bf Theorem 4. Security of entanglement-based key distribution against weakly basis-dependent individual oblivious attacks.}. {\em  Consider an individual attack by Fred, in which Fred applies $U_0^{(i)}$ to the $i$th pair if the basis choice is $a_i=0$ and applies $U_1^{(i)}$ if the basis choice is $a_i=1$, where   ${1\over 4}\|U^{(i)}_0- U^{(i)}_1\|_{\rm sup}^2< \varepsilon$ for each $i$. Then the entanglement-based key distribution protocol is secure, and  secure final key can be extracted from sifted key at the asymptotic rate
\begin{equation}
\label{misalign_rate}
R={\rm Max}\left(1- H_2(\delta)-H_2(\delta + 2f(e\varepsilon),0\right)
\end{equation}	  		                     
where $\delta$  is the bit error rate found in the verification test  and $f(e\varepsilon)$ is defined as in eq.~(\ref{f_define}). } ({\em We assume $\delta + 2f(e\varepsilon)<1/2$}.)
\medskip

In our formulation of Theorem 4, we have chosen to characterize the basis-dependence of the attack in terms of the sup norm distance between the two isometric embeddings $U_0$ and $U_1$ that realize Fred's channels ${\cal E}_0$ and ${\cal E}_1$. It would be more natural to use the intrinsic distance $\|{\cal E}_0-{\cal E}_1\|_{\diamond}$ defined by the ``diamond norm''\cite{diamond}. But the proof of Lemma 3 uses the property that $U_0$ and $U_1$ are close in the sup norm; therefore if we want to reformulate Theorem 4 using the characterization that the channels are close in the diamond norm, we need to show that if two channels are close to one another in the diamond norm, then the dilations of the channels can be chosen to be close in the sup norm. The following lemma, proved in Appendix A, partially solves this problem: 

\medskip
\noindent{\bf Lemma 4.  Similar channels have similar dilations}. {\em Suppose that ${\cal E}_0$ and ${\cal E}_1$ are quantum channels mapping a $d$-dimensional system $S$ to a $d'$-dimensional system $T$, such that $\parallel {\cal E}_0- {\cal E}_1\parallel_{\diamond}<\varepsilon$. Then there are dilations $U_0$ and $U_1$ of the channels (isometric embeddings of $S$ in $TE$, where $E$ is $dd'$-dimensional) such that $\parallel U_0-U_1\parallel_{\rm sup}^2 < d\varepsilon$.
}\medskip

\noindent However, Lemma 4 has the unpleasant property that the dimension $d$ appears in the upper bound on $\parallel U_0-U_1\parallel_{\rm sup}^2$. In principle, the state that Eve delivers to the detector could have arbitrarily high dimension, and Theorem 4 no longer applies if we fix $\varepsilon$ while allowing the dimension to grow without bound. For that reason, we prefer to formulate the statement of Theorem 4 in terms of the sup norm, rather than inferring a bound on the distance between dilations in the sup norm from a bound on the distance between channels in the diamond norm.

\section{Misalignment}

Suppose that Bob is unable to control the orientation of his detector perfectly. When he tries to measure the polarization of his qubit along the $z$-axis, he actually measures along an axis that lies somewhere in a cone around the $z$-axis with opening half-angle $\theta$; similarly when he tries to measure along the $x$-axis, he can only guarantee that his axis is within angle $\theta$ of the desired axis. This scenario is equivalent to one in which Bob's measurement is perfect, but Fred rotates the polarization of the qubit by an angle up to $\theta$ right before the measurement. Furthermore the rotation Fred applies may depend on whether Bob is trying to measure $Z$ or $X$, possibly enhancing the phase error rate relative to the bit error rate. 

Suppose, in addition, that Alice is unable to control the orientation of her source perfectly --- it too might be rotated by an angle up to $\theta$ from the ideal orientation. Equivalently, we may suppose that Alice's source is perfect, but that Fred rotates the qubit slightly (exploiting his knowledge of the basis) immediately after it is emitted by the source. One way to realize such a source is for Alice to prepare a perfect Bell pair $|\phi^+\rangle$ and give half to Fred (who rotates his half); then Alice performs a standard measurement on her half. But a unitary transformation $U$ applied by Fred to his qubit is equivalent to $U^T$ applied to Alice's (where $U^T$ denotes the transpose of $U$); therefore it would make no difference if Alice's qubit were rotated instead of Fred's. Looked at another way, the reason we can replace Fred's rotation by a rotation acting on Alice's qubit is that the source is oblivious --- the emitted state, averaged over the key bits, is maximally mixed, and Fred's attack does not change this property.

In the entanglement-based protocol, then,  the attack in which Fred rotates the orientation of the source and detector is equivalent to an attack in which pairs of qubits are prepared by Eve however she pleases and distributed to Alice and Bob, and then Fred rotates both Alice's and Bob's qubits slightly (by no more than $\theta$) just before standard measurements are performed. Furthermore, we are assuming that Fred's attack is {\em individual} --- the rotation he applies to the $i$th pair is controlled by only the outcome of the flip of the $i$th coin. Therefore, Theorem 4 applies. We can estimate the rate of generation of secure key by calculating the maximum value of 
\begin{equation}
{1\over 4}\|U_0- U_1\|_{\rm sup}^2= \sup_{|\psi\rangle}~ \frac{ {1\over 2}\left(1- {\rm Re}\langle\psi|U_1^{-1}U_0|\psi\rangle\right)}{\langle \psi|\psi \rangle}~,
\end{equation}
where $U_0$ and $U_1$ are unitary transformations applied to the pairs that are consistent with our characterization of the source and detector. 

It is not hard to see that the supremum occurs for $|\psi\rangle$ a maximally entangled state, which, after a suitable choice of basis and phase conventions we may choose to be $|\phi^+\rangle$. Fred applies separate single-qubit rotations to Alice's qubit and to Bob's; acting on $|\phi^+\rangle$, the combined effect of the two is equivalent to a rotation applied to Bob's qubit alone, by an angle no larger than $2\theta$. The overlap $\langle\phi^+|U_1^{-1}U_0|\phi^+\rangle$ is minimized (for $|\theta| \le \pi/4$) if $U_1^{-1}=U_0$; we may choose $U_0$ to be the transformation $I\otimes \tilde U_0$, where $\tilde U_0$ is the single-qubit rotation
\begin{equation}
\tilde U_0=\pmatrix{e^{i\theta} & 0\cr 0& e^{-i\theta}}.
\end{equation}
We find that  
\begin{equation}
\langle\phi^+|I\otimes \tilde U_0^2|\phi^+\rangle= \cos 2\theta~, 
\end{equation}
which implies
\begin{equation}
{1\over 4}\|U_0- U_1\|_{\rm sup}^2=\sin^2\theta~.
\end{equation}
From Lemma 3, then, we find that Fred's attack is $(e\cdot \sin^2\theta + \varepsilon)$-balanced for any positive $\varepsilon$, and we therefore obtain

\medskip
\noindent{\bf Theorem 5. Security of BB84 against individual misalignment of the source and detector}. {\em Suppose that, for each signal, Fred can perform a basis-dependent adjustment of the polarization axes of the source and detector by any angle up to $\theta$. Then the BB84 protocol is secure, and  secure final key can be extracted from sifted key at the asymptotic rate
\begin{equation}
\label{misalign_rate}
R={\rm Max}\left(1- H_2(\delta)-H_2(\delta + 2f(e\cdot \sin^2\theta),0\right)
\end{equation}	  		                     
where $\delta$  is the bit error rate found in the verification test and $f(e\cdot \sin^2\theta)$ is defined as in eq.~(\ref{f_define}).} ({\em We assume  $\delta + 2f(e\cdot \sin^2\theta)<1/2$ and $\theta \le \pi/4)$.})
\medskip

\noindent Thus for $\delta=0$ we obtain a nonzero rate of key generation for $\theta <5.92^\circ$.

We remark again that in the formulation of Theorem 5 the misalignment of the detector or source is assumed to be adversarial, within the angular tolerance specified in our characterization of the device --- Alice and Bob wish to conceal the key from the Eve/Fred alliance. The arguments of Mayers \cite{Mayers96} (for detectors) and Koashi-Preskill \cite{KoashiPreskill02} (for sources) apply to an uncharacterized misalignment that is not adversarial --- Alice and Bob wish to conceal the key from Eve and don't care what Fred knows. In that case, the large potential misalignments do not reduce the key generation rate below that achievable with perfect devices, given a specified bit error rate $\delta$ observed in the test. However, the conclusion of \cite{Mayers96} about security in the case of an uncharacterized detector applies only if the source is perfect, and likewise the conclusion \cite{KoashiPreskill02} about the case of an uncharacterized source applies only if the detector is perfect. In contrast, our analysis applies to the case where both the detector and the source are subject to a characterized misalignment. 

\section{Generic individual flaws in source and detector}

Suppose that the source and detector are both subject to individual flaws that depend weakly on the basis. The source can be modeled as in Sec.~\ref{sec:individual_source}: For each signal to be sent, Fred first prepares a joint state of a qubit $A'$ and a general system $A$. The state that Fred prepares can depend on the basis. Alice then launches the signal by performing a standard measurement on the qubit. If $n$ signals are to be sent, Fred prepares a product state $\bigotimes_{i=1}^n\rho_{a_i}^{(i)}$, where $a_i$ denotes the basis choice for the $i$th signal. Thus we say that Fred's attack on the source is individual. The basis dependence of the source is weak in the sense that $\sqrt{F(\rho^{(i)}_0,\rho^{(i)}_1)}> 1-2\varepsilon_s$ for each $i$. 

We model the detector as follows: Each signal, after Eve's basis-independent attack, is a state of a general system $B$. The signal is received by Fred, who applies a channel that ``squashes'' the signal to a qubit $B'$; Fred's channel may depend on the basis in which Bob will conduct his measurement. Then Bob performs a standard single-qubit measurement on the qubit. Fred's basis-dependent squash can be realized as a basis-dependent isometric embedding of $B$ in $B'E$ where $E$ is a suitable ancilla. If $n$ signals are received by the detector, this transformation can be expressed as the tensor product $\bigotimes_{i=1}^n U^{(i)}_{a_i}$. Thus we say that Fred's attack on the detector is individual. Furthermore the attack depends only weakly on the basis, in the sense that ${1\over 4}\|U^{(i)}_0- U^{(i)}_1\|_{\rm sup}^2< \varepsilon_d$ for each $i$. 

By simultaneously allowing basis-dependent flaws in the source and in the detector, we are going beyond the analysis in Sec.~\ref{sec:individual_source} and Sec.~\ref{sec:individual_pair}. But we may anticipate that, as in those cases considered previously, we can show that the attack is $\Delta$-balanced for small $\Delta$, if $\varepsilon_s$ and $\varepsilon_d$ are small. Indeed, this is the case; for example, if $\varepsilon_s=\varepsilon_d\equiv\varepsilon$ we can show

\medskip
\noindent{\bf Lemma 5}.  {\em Suppose that the source and detector are subject to basis-dependent flaws. The $i$th signal sent by Alice is prepared by performing a standard qubit measurement on half of an entangled state --- this state is $\rho_0^{(i)}$ when the $Z$ basis is declared and $\rho_1^{(i)}$ when the $X$ basis is declared, where $\sqrt{F(\rho_0^{(i)},\rho_1^{(i)})} > 1-2\varepsilon$ for all $i$. The $i$th signal received by the detector is first squashed to a qubit and then a standard measurement is performed. The squash is described by a channel that can be realized by the isometric embedding $U^{(i)}_0$ when the $Z$ basis is declared and by $U^{(i)}_1$ when the $X$ channel is declared, where ${1\over 4}\|U^{(i)}_0- U^{(i)}_1\|_{\rm sup}^2< \varepsilon$ for each $i$. This attack is $\Delta-balanced$ for any $\Delta > 8\sqrt{\varepsilon}+ 4\varepsilon$.}
\medskip

\noindent Lemma 5, together with Theorem 2, provides a proof of security for generic individual flaws in the source and detector that depend sufficiently weakly on the basis. We omit the proof of Lemma 5, which is rather long and unenlightening.

A surprising feature of Lemma 5 is the term scaling like $\sqrt{\varepsilon}$ in our bound on $\Delta$ --- one might reasonably have expected a stronger result, that the attack is $\Delta$-balanced for some $\Delta$ linear in $\varepsilon$. However, we have not succeeded in proving a linear bound.

\section{Tagged signals}
\label{sec:tagged}

Suppose that a fraction $\Delta$ of the qubits emitted by the source are {\em tagged} by Fred. The tag informs Eve which basis was used, so that she can measure the qubit without disturbing it. Eve has no information about the basis used for the untagged qubits (a fraction $1-\Delta$  of the total).

Note that tagged qubits arise in QKD with weak coherent states. The phase of a signal emitted by a coherent light source may be regarded as random if Eve has no information about the phase \cite{Brassard99,EnkFuchs02}, so that the signal state is a mixture of photon number eigenstates. If the source emits more than one photon, we pessimistically assume that Eve stores the extra photons until after the bases are broadcast, and then measures in the proper basis to learn the key bit without introducing any disturbance. Then the tagging probability is $\Delta= p_M/p_D$, where $p_M$ is the probability of emitting a multiphoton, and $p_D$ is the probability that an emitted photon is detected (we pessimistically assume that all of the photons that fail to arrive were emitted as single photons). Arguably we know $p_M$ if we understand our source well, and $p_D$ can be measured. Hence $\Delta$ is a known (or at least knowable) parameter characterizing a practical implementation of quantum key distribution.

We can incorporate tagging into our source model by allowing Fred to append to each qubit emitted by Alice's source an auxiliary qutrit that conveys information about the basis to Eve. For a fraction $\Delta$ of the signals (Fred gets to decide which ones), he sets the value of the qutrit to $|a\rangle$, where $a=0$ indicates the $Z$ basis and $a=1$ indicates the $X$ basis. For the remaining fraction $1-\Delta$ of the qubits sent by Alice, Fred sets the qutrit to $|2\rangle$, passing no basis information to Eve. Eve can read the auxiliary qutrit to learn the basis for each tagged qubit, and so measure the key bit without introducing any disturbance. If each coin that determines the basis choice is a qubit initially prepared in the $X=1$ eigenstate $(|0\rangle +|1\rangle)/\sqrt{2}$, then Fred's attack causes the coin to decohere in the $\{|0\rangle,|1\rangle\}$ basis if the corresponding signal is tagged, but leaves the coin undisturbed if the signal is untagged.

It follows that the attack is $(\Delta/2 + \varepsilon)$-balanced for any positive $\varepsilon$, and we could prove security by applying Theorem 2. But in this case it is possible to prove a stronger result, because we know more about the quantum state of the coins. Suppose that, as in Sec. VII, we apply controlled-NOT gates from the pairs to the coins, transforming $n_{\rm gap}^{(Z)}$ to $\frac{1}{2} \sum_{i=1}^n Z_{{\rm coin}, i}$. The action on the coin of a controlled-NOT gate preserves an $X$-eigenstate. 
Therefore, the probability distribution governing the value of $n_{\rm gap}^{(Z)}$ is the same as the probability distribution governing $\frac{1}{2} \sum_{i=1}^n Z_{{\rm coin}, i}$ in a state of the $n$ coins with the property that $n(1-\Delta)$ of the coins are in eigenstates of $X$ with eigenvalue +1. Hence with high probability $|n_{\rm gap}^{(Z)}|< n\Delta/2 +\varepsilon$ for any positive $\varepsilon$. A similar argument applies to $|n_{\rm gap}^{(X)}|$, and we find that $|n_{\rm gap}|=|n_{\rm gap}^{(Z)}+n_{\rm gap}^{(X)}| < n\Delta +\varepsilon$ for any positive $\varepsilon$. We conclude that secure key can be extracted from sifted key at the asymptotic rate
\begin{equation}
\label{tagged_rate}
R= 1- H_2(\delta) - H_2(\delta + \Delta)~,
\end{equation}
where we have assumed that $\delta +\Delta \le 1/2$. 

Note that to obtain the upper bound on $n_{\rm gap}$, all that we needed was the property that Fred interacts with no more than $n\Delta$ of the coins. Therefore, the argument can be applied more broadly than to the particular tagging model that we have defined above. For example, it applies to a setting where there are flaws in the random number generators used by Alice and Bob to select the basis and the key bits. Suppose that for a fraction $n(1-\Delta)$ of the signals, the basis choice and the key bit are chosen by flipping fair coins, but for a fraction $n\Delta$ of the signals, Fred is free to choose the basis and the key bit however he chooses. In this model, if the source and the detector are perfect otherwise, Fred need not touch $n(1-\Delta)$ of the coins, and secure key can be generated at the rate eq.~(\ref{tagged_rate}). (In this estimate of the rate, however, we have continued to assume that the qubits selected for the verification test are a fair sample, and so provide an accurate estimate of the error rate for the key generating pairs. The argument can be extended further to cover the case where Fred is permitted to select a small portion of the test set, by adjusting the estimate of the error rate to take into account the bias in the test.)

With a more sophisticated argument we can obtain a higher rate of secure key generation than eq.~(\ref{tagged_rate}). After correcting errors in the sifted key (sacrificing a fraction $H_2(\delta)$ of the key, asymptotically) we imagine executing privacy amplification on two different strings, the sifted key bits arising from the tagged qubits and the sifted key bits arising from the untagged qubits. Since the privacy amplification scheme described in Sec.~III is linear (the private key can be computed by applying the $C_2$ parity check matrix to the sifted key after error correction), the key obtained is the bitwise XOR 
\begin{equation}
s_{\rm untagged}\oplus s_{\rm tagged}
\end{equation}
of keys that could be obtained from the tagged and untagged bits separately. If $s_{\rm untagged}$ is private and random, then it doesn't matter if Eve knows everything about $s_{\rm tagged}$ -- the sum is still private and random.

Therefore we ask if privacy amplification is successful applied to the untagged bits alone. Under the worst case assumption that the bit error rate is zero for tagged qubits, the overall bit error rate $\tilde \delta$ is related to the bit error rate $\tilde\delta_{\rm untagged}$ for the untagged qubits by
\begin{equation}
\tilde\delta = (1-\Delta)\tilde\delta_{\rm untagged}~.
\end{equation}
Since the bit errors and phase errors are related by symmetry for the untagged qubits, the phase error rate $\tilde\delta_{p,{\rm untagged}}$ for the untagged qubits
satisfies
\begin{equation}
\tilde\delta_{p,{\rm untagged}} < \tilde\delta_{\rm untagged} + \varepsilon = {\tilde\delta\over 1-\Delta} +\varepsilon
\end{equation}
with high probability.
Since the error rate $\delta$ observed in the test provides a good estimate of $\tilde\delta$ for the key generating pairs, we conclude that 
\begin{equation}
\tilde\delta_{p,{\rm untagged}} < {\delta\over 1-\Delta}+\varepsilon'~.
\end{equation}
with high probability, for any positive $\varepsilon'$ and sufficiently large $n$.
If there are $n$ bits of sifted key, then $(1-\Delta)n$ of these bits come from untagged qubits, and (since bit errors are already corrected) we can extract a private key by sacrificing a fraction $H_2(\tilde\delta_{p,{\rm untagged}}+\varepsilon'')$ of these for privacy amplification. Thus we have proved:

\medskip
\noindent{\bf Theorem 6. Security of BB84 against tagging}. {\em Suppose that Fred interacts with only $n\Delta$ of the $n$ coins that determine the basis used by Alice and Bob. Then the BB84 protocol is secure, and secure final key can be extracted from sifted key at the asymptotic rate
\begin{equation}
\label{wcs_rate}
R={\rm Max}\left((1-\Delta) - H_2(\delta) - (1-\Delta)H_2\left({\delta\over 1-\Delta}\right),0\right)
\end{equation}	  		                     
where $\delta$  is the bit error rate found in the verification test (assuming $\delta/(1-\Delta)<1/2$). In particular, this rate of key generation is achievable, assuming that the source and the detector are perfect otherwise, if Fred reveals the basis to Eve for $n\Delta$ of the signals, or if Fred chooses the basis and key bits for $n\Delta$ of the signals. }
\medskip

\noindent 

In the case where the source emits weak coherent states with random phases, a rate of key generation similar to eq.~(\ref{wcs_rate}) was established by Inamori, L\"utkenhaus, and Mayers (ILM) \cite{ILM01}.  
Actually, the rate quoted by ILM is below $R$ in Eq.~(\ref{wcs_rate}) --- in their Eq.~(18) the argument of $H_2$ in the last term is $2\delta/(1-\Delta)$ rather than $\delta/(1-\Delta)$. However, we believe that their argument can be refined to match the rate Eq.~(\ref{wcs_rate}). With that refinement the ILM result is stronger in a sense than what we have derived here, as it applies to the case of a general uncharacterized detector. 

Theorem 6 can be applied if there is loss in the quantum channel connecting Alice and Bob and/or if Bob's detector has imperfect efficiency, provided that the loss is basis-independent. For example, suppose that each signal emitted by Alice's source is a phase-randomized weak coherent state with mean photon number $\mu\ll 1$, so that the signal is a single photon with probability $p_1= \mu +O\left(\mu^2\right)$, and more than one photon with probability $p_M=\frac{1}{2}\mu^2+O\left(\mu^3\right)$. We can describe these signals by imagining a source that never emits multiple photons, followed by a basis-dependent attack by Fred in which Fred interacts with a fraction $p_M$ of all the coins. Now suppose that Eve's attack can be modeled by a basis-independent lossy channel, such that a fraction $\eta$ of all the nonvacuum signals are detected. (Here by ``basis-independent'' we mean that Eve's attack has no {\it a priori} dependence on the basis, though of course Eve can exploit the multiphotons to acquire some information about the basis; the important thing is that Eve can launch her attack without interacting with the coins.) Then a fraction $p_D=\eta\left(\mu +O(\mu^2)\right)$ of all the signals are detected, and of the coins associated with detected signals, Fred interacts with at most a fraction
\begin{equation}
\Delta= p_M/p_D= \frac{1}{2\eta}\left(\mu + O\left(\mu^2\right)\right)~.
\end{equation}
Sifted key is generated at the rate 
\begin{equation}
\frac{1}{2}\nu p_D\approx \frac{1}{2}\nu \eta \mu \approx  \nu \eta^2\Delta~,
\end{equation}
where $\nu$ is the repetition frequency of the source. 
Therefore, if $\Delta$ (and hence also the rate $R$ of generation of final key from sifted key) is held fixed as $\eta$ gets small, then the overall key generation rate $\frac{1}{2}\nu p_D R$ is $O(\eta^2)$, as ILM observed \cite{ILM01}. This scaling of the rate with $\eta$ holds approximately as long as dark counts in the detector are not too important, so that the bit error rate $\delta$ is roughly independent of $\eta$.  In some current implementations of quantum key distribution using weak coherent states transmitted through optical fibers, dark counts are relatively unimportant, and our analysis of security is applicable, up to a range of approximately 20 km. 

Theorem 6 applies if Fred tags any $n\Delta$ of the signals. But it does not apply to a {\em coherent superposition} of such attacks. Suppose, for example, that in the entanglement-based protocol, Fred's attack on the pairs and the coins produces a state
\begin{equation}
\label{tagged_sup}
|\Psi\rangle = \sum_{S:|S|\le n\Delta} a_S|\Psi_S\rangle~;
\end{equation}
here the sum is over subsets $S$ that contain no more than $n\Delta$ of the $n$ pairs, and $|\Psi_S\rangle$ is the state resulting from tagging the pairs in the set $S$. Although Theorem 6 does not apply to a general superposition of tagged states as in eq.~(\ref{tagged_sup}), Theorem 2 {\em does} apply to this case. After we trace out Fred's labeling qutrits, the state of the coin can be realized as an ensemble of states, where for each state in the ensemble at least $n(1-\Delta)$ of the coins are $X=1$ eigenstates and the rest are mixtures of $Z$ eigenstates. Therefore, if the coins are all measured in the $X$ basis, with high probability the number of coins for which the outcome $X=-1$ is found will be less than $n(\Delta/2 +\varepsilon)$. Thus the attack is $(\Delta/2 + \varepsilon)$-balanced for any positive $\varepsilon$ and sufficiently large $n$, and it follows from Theorem 2 that secure key can be generated at the corresponding rate (a lower rate than found in Theorem 6). 

In particular, then, Theorem 2 can be applied to a general source that emits signals that are sufficiently close to perfect single photon pulses, even if the multiphotons occur with nonrandom phases. Unfortunately, though, our arguments do not allow us to address the case where the source emits weak coherent states with nonrandom phases ---  in that case the states are dominated by the amplitude to emit the vacuum state, and the tagging model we have analyzed here does not apply. This difficulty seems to be more than a mere shortcoming of our model; the deeper problem is that weak coherent states with nonrandom phases leak a significant amount of basis information, which may compromise security.

As for all of the cases that we consider in this paper, the crux of our analysis of tagging is a bound on the phase error rate $\tilde\delta_p$ of the key generating pairs that holds {\em with high probability} --- it does not suffice for $\tilde\delta_p$ to be bounded after {\em averaging} over Fred's strategy. Therefore, our security proof need not apply for a highly correlated basis-dependent attack on the signals, even if the bit error rate $\tilde\delta$ and phase error rate $\tilde\delta_p$ resulting from the attack have mean values that are nearly equal.\footnote{We thank Dominic Mayers for a helpful discussion of this point.} 

For example, suppose that with a small probability $r$, Fred tells Eve the basis for {\em every} signal, while with probability $1-r$, Fred tells Eve nothing. Then on average the disparity between the bit error rate and the phase error rate is small. However, with a fixed probability $r$ that does not depend on the key length, Eve can learn the whole key. Therefore, the quantum key distribution protocol is insecure for a source of this type.

\section{Trojan pony}
\label{sec:pony}
Suppose that the detector is not perfectly efficient. A fraction $\Delta$ of the signals that enter the detector fail to trigger it, resulting in no recorded outcome. Suppose further that Fred, who knows Bob's basis, controls whether the detector fires or not, subject to the constraint  that only a fraction $\Delta$ of the detection events can be eliminated. Note that the parameter $\Delta$ can be measured in the protocol.

Fred can use his power to disguise Eve's attack, enhancing the detection rate when Bob measures in the same basis as Eve did and suppressing the detection rate when Bob measures in a different basis than Eve's. This is a limited version of the ``Trojan horse'' attack \cite{Lo01} --- we call it the ``Trojan pony.'' As we remarked in Sec.~\ref{sec:real}, one version of the Trojan pony attack can be launched if Bob's detector is configured as a polarization beam splitter that directs the signals to a pair of threshold detectors; Eve can ensure that the detector fails to register a conclusive result by flooding it with many photons. We will analyze this attack in a different setting, in which Bob's detector receives qubits rather than bosonic modes.

In the EDP setting, we allow Fred to eliminate a fraction $\Delta$ of the pairs (corresponding to the qubits for which he ``turns off'' Bob's detector). In the worst case, every pair that he eliminates has a bit error and no phase error. Before any pairs were eliminated, the error rate was essentially the same in both bases --- call this rate $p$. After eliminating the undetected pairs, the error rates are
\begin{equation}
\label{delta_trojan}
\tilde\delta \approx{ p-\Delta\over 1-\Delta}~,\quad 
\tilde \delta_p\approx{p\over 1-\Delta}~
\end{equation}
(assuming $\Delta\le p\le 1-\Delta$). Note that, for ease of presentation, we have not included the $\varepsilon$'s in eq.~(\ref{delta_trojan}); instead we have used the symbol $\approx$ to indicate relations that are satisfied to arbitrarily good accuracy with high probability asymptotically. Eliminating $p$ we find
\begin{equation}
\tilde \delta_p\approx \tilde\delta+ {\Delta\over 1-\Delta}~,
\end{equation}
and, since the error rate $\delta$ measured in the test provides a reliable estimate of $\tilde\delta$, we infer that final key can be generated from sifted key at the achievable rate 
\begin{equation}
\label{pony_rate}
R=1-H_2(\delta)- H_2\left(\delta + {\Delta\over 1-\Delta}\right)~,
\end{equation}
where we have assumed that
\begin{equation}
\delta + {\Delta\over 1-\Delta}\le 1/2~.
\end{equation}

We can use similar reasoning if the detector efficiency is low, but we trust that most of the instances where the detector fails to fire are chosen at random, and only a small percentage of all the detector failures are due to Fred's intervention. In the absence of other imperfections, random misfires merely reduce the number of sifted key bits, but without breaking the symmetry between the bases. Eq.~(\ref{pony_rate}) still applies if a fraction $f$ of detection events are removed by random errors, and a fraction $\Delta$ of the remaining events are removed adversarially, resulting in an overall efficiency $\eta=(1-f)(1-\Delta)$. Thus we have proved 

\medskip
\noindent{\bf Theorem 7. Security of BB84 against basis-dependent detector efficiency}. {\em Suppose that of the signals that arrive at Bob's detector, a fraction $f$ chosen at random are removed, and of those that remain a fraction $\Delta$ chosen adversarially by Fred are also removed, so that the overall efficiency of the detector is $\eta=(1-f)(1-\Delta)$. Then the BB84 protocol is secure, and secure final key can be extracted from the (detected) sifted key at the asymptotic rate
\begin{equation}
R={\rm Max}\left(1-H_2(\delta)- H_2\left(\delta + {\Delta\over 1-\Delta}\right),0\right)
\end{equation}	  		                     
where $\delta$  is the bit error rate found in the verification test (assuming $\delta+ \Delta/(1-\Delta)<1/2$).}
\medskip

\noindent Note that we can measure the efficiency $\eta$ in the protocol, but can determine $\Delta$ only by acquiring a good understanding of the vulnerability of the detector to tampering. In fact, in current implementations the typical efficiency for detection of single photons at telecommunication wavelengths is about $15\%$ \cite{Gisin02}. Theorem 7 can also be applied to the case where basis-dependent losses occur in the quantum channel connecting Alice and Bob, with $\Delta$ parametrizing the basis dependence.

ILM \cite{ILM01,Norbert98} discussed the specific type of Trojan pony attack in which Eve floods Bob's polarization beam splitter with many photons of the same polarization, generating a ``double click'' in Bob's two photon detectors when he tries to measure the polarization in the conjugate basis. For this case they proposed that Bob choose his key bit randomly each time he encounters a double click event. Security of this scheme is ensured by the result of Mayers \cite{Mayers96}: the POVM that assigns a random outcome to the ``double-click'' subspace is a possible measurement that Fred could arrange, and Mayers proved security for an arbitrary detector POVM. If double clicks occur a fraction $\Delta$ of the time, and the bit error rate is $\delta$ when single clicks occur, then the overall error rate under the ILM prescription will be $(1-\Delta)\delta +\Delta/2$, resulting in a key generation rate
\begin{equation}
\label{ILM_pony_rate}
R= 1 - 2 H_2\Big((1-\Delta)\delta + \Delta/2\Big)~.
\end{equation}
The rate is further enhanced by the factor $(1-\Delta)^{-1}$ relative to Eq.~(\ref{pony_rate}), since all detection events, including the double clicks, contribute to the sifted key. Thus the achievable rate established by ILM exceeds the rate we have derived, except for relatively large $\Delta$ and relatively small $\delta$. However, the two results cannot be compared directly, because they apply to two different models of the adversary. The ILM result Eq.~(\ref{ILM_pony_rate}) applies to a particular Trojan pony attack that can be launched by Eve if the Bob/Fred POVM has suitable properties; it provides a condition for Eve (but not Fred) to have negligible information about the key. Eq.~(\ref{pony_rate}) is the rate at which key can be extracted under a Trojan pony attack in which Fred receives qubits, and can prevent some of the qubits from registering in Bob's detector. But in this case the key is kept secret not just from Eve but from the Eve/Fred alliance.

A different type of issue relating to detector efficiency arises if the detector failures occur at different rates when measuring in the $X$ and $Z$ bases, but are otherwise randomly distributed. The bias in the detector efficiency breaks the symmetry between the bases, but a simple variant of the Shor-Preskill argument still applies. In the entanglement distillation picture, we may imagine that Alice and Bob at first share many noisy pairs; furthermore, after a random permutation unknown to the adversary is applied, the pairs are symmetrized so that all have the same marginal density operator. Then some of the pairs are removed from the sample by a random process. Though the probability of removal may depend on the basis used to generate the key bit, Alice and Bob can still infer the phase error rate from the bit error rate if they conduct a refined data analysis \cite{LoChauArdehali00}, measuring separate error rates $\delta_X$ and $\delta_Z$ for the $X$ and $Z$ bases respectively. If a fraction $p_X$ of the sifted key bits are generated in the $X$ basis and a fraction $p_Z$ in the $Z$ basis (where $p_X+p_Z=1$), so that the bit error rate is $\delta=p_X\delta_X+p_Z\delta_Z$, then the phase error rate to insert in Eq.~(\ref{key_rate}) becomes $\delta_p=p_X\delta_Z+p_Z\delta_X$.

\section{Conclusions}

We have shown that the BB84 quantum key distribution protocol is secure when the source and/or detector are subject to small errors that are controlled by an adversary who knows the basis used by Alice and Bob. We have formulated a method for estimating the key generation rate in the presence of such errors, and we have applied the method to various model sources and detectors. Our results are complementary to earlier proofs of security \cite{Mayers96,KoashiPreskill02} that apply to flaws in the apparatus that may be large but are nonadversarial; furthermore, our results unlike those of \cite{Mayers96,KoashiPreskill02} apply when both the source and the detector have small basis-dependent flaws, as will be the case in typical real-world implementations of quantum key distribution. We have argued that the security holes of real sources and detectors can be usefully investigated within our framework, and we expect that the methods we have developed will find further applications. 

However, the model sources and detectors to which our analysis applies are not completely general. In our model of the source, each signal is launched by preparing an entangled state of a {\em qubit} and a general system, followed by an ideal measurement of the qubit. To establish security, we require that the entangled state depend only weakly on the basis used in the protocol. With this model, we are unable to treat, for example, the case where the source emits weak coherent states with {\em nonrandom} phases. Likewise, we model the detector as a quantum channel followed by an ideal measurement of a qubit, and to establish security we require that the quantum channel depend only weakly on the basis. In particular, we are unable to treat the case where the signals received by the detector reside in a Hilbert space of arbitrarily high dimension.

Various other issues regarding the security of BB84 and other quantum key distribution protocols have not been addressed here.  We have not considered how to characterize devices reliably using testing equipment that is itself untrustworthy (as in \cite{MayersYao98}).  We have not discussed how to improve the rate of key generation beyond the rate in Eq.~(\ref{key_rate}) through privacy amplification schemes that use two-way communication between Alice and Bob \cite{GottesmanLo01}. Finally, our security analysis applies to the asymptotic limit of an infinite key  --- we have not analyzed the practical aspects of error correction and privacy amplification in the case of finite key length.

\acknowledgments
We thank Michael Ben-Or, Jim Harrington, Masato Koashi, Dominic Mayers, and Peter Shor for helpful discussions. This work has been supported in part by: the Department of Energy under Grant No. DE-FG03-92-ER40701,  the National Science Foundation under Grant No. EIA-0086038,  the Caltech MURI Center for Quantum Networks under ARO Grant No. DAAD19-00-1-0374, the Clay Mathematics Institute, Canadian NSERC, Canada Research Chairs Program, Canadian Foundation for Innovation, Ontario Innovation Trust, Premier's Research Excellence Award, Canadian Institute for Photonics Innovation, MagiQ Technologies, Inc., and the German Research Council (DFG) under the Emmy-Noether Programme. 

\section*{Appendix A. Similar channels have similar dilations}
Here we will prove:

\medskip
\noindent{\bf Lemma 4.  Similar channels have similar dilations}. {\em Suppose that ${\cal E}_0$ and ${\cal E}_1$ are quantum channels mapping a $d$-dimensional system $S$ to a $d'$-dimensional system $T$, such that $\parallel {\cal E}_0- {\cal E}_1\parallel_{\diamond}<\varepsilon$. Then there are dilations $U_0$ and $U_1$ of the channels (isometric embeddings of $S$ in $TE$, where $E$ is $dd'$-dimensional) such that $\parallel U_0-U_1\parallel_{\rm sup}^2 < d\varepsilon$.
}\medskip

It is convenient to characterize a quantum channel ${\cal E}$ mapping system $S$ to system $T$ by considering the action of $I\otimes {\cal E}$ on a reference system $R$ and the system $S$, where ${\rm dim}~ R= {\rm dim}~S\equiv d$. Let
\begin{equation}
|\tilde \Phi\rangle = \sum _i |i\rangle_R\otimes |i\rangle_S
\end{equation}
denote an unconventionally normalized maximally entangled pure state on $RS$, satisfying $\langle\tilde\Phi|\tilde\Phi\rangle=d$. We may define
\begin{equation}
\tilde \rho=I\otimes {\cal E}\big(|\tilde\Phi\rangle\langle \tilde\Phi|\big)~,
\end{equation}
where $\tilde \rho$ is an unconventionally normalized density operator on $RT$, satisfying ${\rm tr}~\tilde\rho=d$. The action of ${\cal E}$ on a pure state $|\varphi\rangle=\sum_i a_i|i\rangle$ on $S$ can then be expressed as
\begin{equation}
{\cal E}\big(|\varphi\rangle\langle\varphi|\big)= \langle\varphi^*|\tilde\rho|\varphi^*\rangle
\end{equation}
where $|\varphi^*\rangle=\sum_i a_i^* |i\rangle$ is the ``index state'' on $R$ corresponding to $|\varphi\rangle$. If we introduce an additional system $E$ (the ``environment,'' of dimension $dd'$), we can construct a purification $|\tilde\Phi'\rangle$ of $\tilde\rho$ on $RTE$ such that $\langle \tilde\Phi'|\tilde\Phi'\rangle=d$. This purification defines a ``dilation'' $U$ of the channel ${\cal E}$ that realizes ${\cal E}$ as an isometric embedding of $S$ in $TE$. The action of the dilation on the pure state $|\varphi\rangle$ is
\begin{equation}
|\varphi\rangle \to U|\varphi\rangle=\langle \varphi^*|\tilde\Phi'\rangle ~.
\end{equation}

Now suppose that ${\cal E}_0$ and ${\cal E}_1$ are two channels acting on $S$, satisfying the inequality 
\begin{equation}
\| {\cal E}_0 - {\cal E}_1\|_\diamond < \varepsilon~,
\end{equation}
and that $\tilde \rho_0$ and $\tilde \rho_1$ are the corresponding states obtained from the action of $I\otimes {\cal E}_0$ and $I\otimes {\cal E}_1$ on $|\tilde \Phi\rangle$. The diamond norm \cite{diamond} is defined by
\begin{equation}
\|{\cal E}\|_\diamond \equiv \sup_{X\ne 0}\frac{\| I\otimes {\cal E}(X)\|_{\rm tr}}
{\| X \|_{\rm tr}}~;
\end{equation}
since $\langle \tilde \Phi|\tilde \Phi\rangle=d$,
it follows that the trace distance between $\tilde\rho_0$ and $\tilde\rho_1$ satisfies
\begin{equation}
\parallel \tilde \rho_0 - \tilde\rho_1\parallel_{\rm tr} ~< d\varepsilon~.
\end{equation}
Since, for conventionally normalized density operators, the fidelity and trace distance are related by
\begin{equation}
\sqrt{F(\rho_0,\rho_1)} \ge 1 - {1\over 2}\parallel \rho_0 -\rho_1\parallel_{\rm tr}~,
\end{equation}
it follows \cite{uhlmann76,jozsa94} that $\tilde \rho_0$ and $\tilde \rho_1$ have purifications $|\tilde \Phi_0'\rangle$ and $|\tilde \Phi_1'\rangle$ on $RSE$ with norm $\sqrt{d}$ and overlap satisfying
\begin{equation}
\label{purification_overlap}
{\rm Re}~ \langle\tilde \Phi_1'|\tilde \Phi_0'\rangle > d\left(1-{\varepsilon\over 2}\right)~.
\end{equation}

This large overlap of the purifications $|\tilde \Phi_0'\rangle$ and $|\tilde \Phi_1'\rangle$ implies that the corresponding dilations $U_0$ and $U_1$ of the channels are close to one another in the sup norm. 
Given any $|\varphi\rangle$ on $S$, we may regard it as one element of a basis $\{|\varphi_i\rangle\}$ for $S$. Then eq.~(\ref{purification_overlap}) may be rewritten as
\begin{eqnarray}
&&{\rm Re}~ \sum_{i=1}^d \langle \varphi_i|U_1^\dagger U_0|\varphi_i\rangle\nonumber\\
&&={\rm Re}~ \sum_{i=1}^d\langle\tilde \Phi_1'|\varphi_i^*\rangle\langle \varphi_i^*|\tilde \Phi_0'\rangle > d\left(1-{\varepsilon\over 2}\right)~.
\end{eqnarray}
But each of the $d$ terms in the sum is no larger than 1, and since the sum is greater than $d-d\varepsilon/2$, each term must be greater than $d-d\varepsilon/2 -(d-1)= 1-d\varepsilon/2$. We conclude, then, that for any pure state $|\varphi\rangle$ on $S$, 
\begin{equation}
\label{overlap_bound}
{\rm Re}~ \langle \varphi|U_1^\dagger U_0|\varphi\rangle > 1-{d\varepsilon\over 2}~.
\end{equation}
Therefore, for any $|\varphi\rangle$
\begin{equation}
\parallel (U_0-U_1)|\varphi\rangle\parallel^2 <d\varepsilon~,
\end{equation}
and hence
\begin{equation}
\parallel U_0-U_1\parallel_{\rm sup}^2 < d\varepsilon~.
\end{equation}
This proves Lemma 4.

\end{multicols}

\begin{thebibliography}{10}

\bibitem{BennettBrassard84}
C.~H. Bennett and G. Brassard, ``Quantum cryptography: Public key distribution and coin tossing,'' in {\em Proceedings of IEEE International
  Conference on Computers, Systems and 
Signal Processing, Bangalore, India}
  (IEEE, New York, 1984), pp. 175--179.

\bibitem{Mayers96}
D.~Mayers, ``Quantum key distribution and string oblivious transfer in noisy channels,''
in {\em Advances in 
Cryptography---Proceedings of Crypto'96} (Springer-Verlag, New York, 1996),
pp. 343-357; ``Unconditional security in quantum cryptography,'' 
J.~Assoc.~Comp.~Mach. {\bf 48}, 351 (2001),
arXiv:quant-ph/9802025.

\bibitem{LoChau99}
H.-K. Lo and H.~F. Chau,
``Unconditional security of quantum key distribution over arbitrarily
long distances,''
{Science} {\bf 283}, 2050--2056 (1999), arXiv:quant-ph/9803006. 


\bibitem{BBBMR00}
E.~Biham, M.~Boyer, P.~O.~Boykin, T.~Mor, and V.~Roychowdhury, ``A proof of the security of quantum key distribution,''
in {\em Proceedings of the 32nd Annual ACM Symposium on Theory of
Computing} (ACM Press, New York, 2000), pp. 715--724, arXiv:quant-ph/9912053.


\bibitem{ShorPreskill00}
P.~W. Shor and J. Preskill, ``Simple proof of security of the BB84 quantum key distribution protocol,'' Phys.~Rev.~Lett. {\bf 85}, 441--444 (2000), arXiv:quant-ph/0003004.


\bibitem{KoashiPreskill02}
M.~Koashi and J.~Preskill, 
``Secure quantum key distribution with an uncharacterized source,'' Phys. Rev. Lett. {\bf 90}, 057902 (2003),
arXiv:quant-ph/0208155 (2002).

\bibitem{MayersYao98}
D. Mayers and A. Yao, ``Quantum cryptography with imperfect apparatus,'' arXiv:quant-ph/9809039 (1998); D. Mayers and A. Yao, ``Self testing quantum apparatus,'' arXiv:quant-ph/0307205.

\bibitem{ILM01}
H.~Inamori, N.~L\"utkenhaus and D.~Mayers, ``Unconditional security of practical quantum key distribution,'' arXiv:quant-ph/0107017 (2001).

\bibitem{Slutsky98}
B.~A. Slutsky, R. Rao, P.-C. Sun, and Y. Fainman, ``Security of quantum cryptography against individual attacks,'' Phys. Rev. A {\bf 57}, 2383--2398 (1998).

\bibitem{Norbert99}
N. L\"utkenhaus, ``Security against individual attacks for realistic quantum key distribution,'' Phys.~Rev. A {\bf 61}, 052304 (2000), arXiv:quant-ph/9910093. 

\bibitem{Brassard99} 
G. Brassard, N. L\"utkenhaus, T. Mor, and B.~C. Sanders, ``Security aspects of practical quantum cryptography,'' Phys.~Rev.~Lett. {\bf 85}, 1330--1333 (2000), arXiv:quant-ph/9911054.

\bibitem{Felix01}
S. Felix, N. Gisin, A. Stefanov, H. Zbinden, ``Faint laser quantum key distribution: Eavesdropping exploiting multiphoton pulses,'' J. Mod. Opt. {\bf 48}, 2009 (2001), arXiv:quant-ph/0102062.

\bibitem{Gilbert00} 
G. Gilbert and M. Hamrick, ``Practical quantum cryptography: a comprehensive analysis (part one),'' arXiv:quant-ph/0009027 (2000).

\bibitem{Gilbert01}
G. Gilbert and M. Hamrick, ``Secrecy, computational loads and rates in practical quantum cryptography,'' Algorithmica {\bf 34}, 314-339 (2002), arXiv:quant-ph/0106043 (2001).

\bibitem{GottesmanLo01}
D.~Gottesman and H.-K.~Lo, 
``Proof of security of quantum key distribution with two-way classical communications,'' IEEE Trans. Information Theory {\bf 49}, 457 (2003),
arXiv:quant-ph/0105121 (2001).

\bibitem{Ben-Or}
M. Ben-Or,``Simple security proof for quantum key distribution,'' online presentation available at http://www.msri.org/publications/ln/msri/2002/qip/ben-or/1/index.html (2002).

\bibitem{Bennett95}
C. H. Bennett, G. Brassard, S. Popescu, B. Schumacher, J. A. Smolin, and W. K. Wootters, ``Purification of noisy entanglement and faithful teleportation via noisy channels,'' Phys. Rev. Lett. {\bf 76}, 722-725 (1996), arXiv:quant-ph/9511027. Erratum: Phys. Rev. Lett. {\bf 78}, 2031 (1997).

\bibitem{Deutsch95}
D. Deutsch, A. Ekert, R. Jozsa, C. Macchiavello, S. Popescu, and A. Sanpera, ``Quantum privacy amplification and the security of quantum cryptography over noisy channels,'' Phys. Rev. Lett. {\bf 77}, 2818-2821 (1996), arXiv.org:quant-ph/9604039. Erratum: Phys. Rev. Lett. {\bf 80}, 2022 (1998). 


\bibitem{BDSW96}
C.~H. Bennett, D.~P. DiVincenzo, J.~A. Smolin and W.~K. Wootters,
``Mixed state entanglement and quantum error correction,''
Phys.~Rev. A {\bf 54}, 3824--3851 (1996),
arXiv:quant-ph/9604024.

\bibitem{CalderbankShor96}
A.~R. Calderbank and P.~W. Shor, 
``Good quantum error correcting codes exist,''
Phys. Rev. A {\bf 54}, 1098--1105 (1996),
 arXiv:quant-ph/9512032.

\bibitem{Steane96} A.~M. Steane, 
``Multiple particle interference and quantum error correction,'' 
{Proc.~Roy.~Soc.~Lond. A}
{\bf 452}, 2551--2577 (1996),
 arXiv:quant-ph/9601029.

\bibitem{GottesmanPreskill01}
D.~Gottesman and J.~Preskill, ``Secure quantum key distribution using squeezed states,'' Phys.~Rev. A {\bf 63}, 022309 (2001), arXiv:quant-ph/0008046.

\bibitem{Hamada03} M. Hamada, ``Reliability of Calderbank-Shor-Steane codes and the security of quantum key distribution,'' arXiv:quant-ph/0311003 (2003).

\bibitem{spielman} D.~A. Spielman, ``Linear-time encodable and decodable error-correcting codes, IEEE Trans. Information Theory {\bf 42}, 1723--1731 (1996).

\bibitem{LoChauArdehali00}
H.-K.~Lo, H.~F. Chau, and M.~Ardehali, 
``Efficient quantum key distribution scheme and proof of its unconditional security,'' arXiv:quant-ph/0011056 (2000).

\bibitem{GottesmanChuang} D. Gottesman and I.~L. Chuang, ``Quantum digital signatures,'' arXiv:quant-ph/0205032 (2001).
\bibitem{Norbert98}
N. L\"utkenhaus, ``Estimates for practical quantum cryptography,'' Phys. Rev. A {\bf 59} 3301--3319 (1999), arXiv:quant-ph/9806008.

\bibitem{Lo01}
H.-K.~Lo, ``Proof of unconditional security of six-state quantum key distribution scheme,'' Quant. Info. Comp. {\bf 1}, 81--94 (2001), arXiv:quant-ph/0102138.

\bibitem{uhlmann76}
A. Uhlmann, ``The `transition probability' in the state space of a ${}^*$-algebra,'' Reports on Mathematical Physics {\bf 9}, 273--279 (1976).

\bibitem{jozsa94}
R. Jozsa, ``Fidelity for mixed quantum states,'' J. Mod. Opt. {\bf 41}, 2315-2323 (1994).

\bibitem{diamond}
D. Aharonov, A, Kitaev, and N. Nisan, ``Quantum circuits with mixed states,'' in {\em Proceedings of the Thirtieth Annual ACM Symposium on Theory of Computing (STOC)} (ACM Press, New York, 1998), pp. 20-30, arXiv:quant-ph/9806029.

\bibitem{EnkFuchs02}
S.~J.~van Enk and C.~A.~Fuchs, ``The quantum state of a laser field,'' Quant. Info. Comp. {\bf 2}, 151--165 (2002), arXiv:quant-ph/0111157.

\bibitem{Gisin02}
 N.~Gisin, G.~Ribordy, W.~Tittel, and H.~Zbinden, ``Quantum cryptography,'' Rev. Mod. Phys. {\bf 74}, 145-195 (2002) arXiv:quant-ph/0101098


 
\end{thebibliography}
\end{document}